\documentclass[12pt,preprint]{aastex}

\shorttitle{Accretion Disks around Young Stars} 
\shortauthors{Jayawardhana et al.}

\begin{document}


\title{Accretion Disks around Young Stars:\\ 
Lifetimes, Disk Locking and Variability}

\author{Ray Jayawardhana, Jaime Coffey, Alexander Scholz,\\ 
Alexis Brandeker \& Marten H. van Kerkwijk}
\affil{Department of Astronomy \& Astrophysics, University of Toronto,
    Toronto, Ontario M5S3H8, Canada}
\email{rayjay@astro.utoronto.ca}

\begin{abstract}
We report the findings of a comprehensive study of disk accretion
and related phenomena in four of the nearest young stellar
associations spanning 6-30 million years in age, an epoch that
may coincide with the late stages of planet formation. We have obtained 
$\sim$650 multi-epoch high-resolution optical spectra
of 100 low-mass stars that are likely members of the $\eta$
Chamaeleontis ($\sim$6 Myr), TW Hydrae ($\sim$8 Myr), $\beta$
Pictoris ($\sim$12 Myr) and Tucanae-Horologium ($\sim$30 Myr)
groups. Our data were collected over 12 nights between 2004 December
- 2005 July on the Magellan Clay 6.5m telescope. Based on H$\alpha$
line profiles, along with a variety of other emission lines, we find
clear evidence of on-going accretion in three out of 11 $\eta$ Cha stars 
and two out of 32 TW Hydrae members. None of the 57 $\beta$ Pic or
Tuc-Hor members shows measurable signs of accretion. Together, these results
imply significant evolution of the disk accretion process within the first
several Myr of a low-mass star's life. While a few disks can continue to
accrete for up to $\sim$10 Myr, our findings suggest that disks accreting
for beyond that timescale are rather rare. This result provides an indirect
constraint on the timescale for gas dissipation in inner disks and, in turn,
on gas giant planet formation. All accretors in our sample are
slow rotators, whereas non-accretors cover a large range in rotational
velocities. This may hint at rotational braking by disks at ages up to 
$\sim$8 Myr. Our multi-epoch spectra confirm that emission-line 
variability is common even in somewhat older T Tauri stars, among which 
accretors tend to show particularly strong variations. Thus, 
our results indicate that accretion and wind activity undergo 
significant and sustained variations throughout the lifetime of 
accretion disks. 
\end{abstract}


\section{Introduction}
Since planets are born in circumstellar disks, studies of disk evolution 
can provide useful constraints on the timescale for planet formation. The 
recently identified young stellar associations, located within 100 pc of 
the Sun and spanning the critical age range of $\sim$5 to 30 million years 
(Myr), constitute superb targets for such investigations. The growing
consensus, primarily from infrared observations of these groups and other 
young clusters, is that (at least) the inner disks are cleared of small 
dust grains within the first 6--10 Myr \citep[e.g.][]{hll01,jhf99b,hja05}.
Unfortunately, little is known about the timescale for the dissipation 
of gas, which accounts for $\sim$99\% of the disk mass and is crucial for 
building giant planets. That is because cold molecular gas, especially 
H$_2$, is hard to detect \citep[e.g.,][]{r02}. It is possible to 
determine the presence of gas in the inner disk, albeit indirectly, from 
evidence of gas being accreted from the inner disk edge on to the central
star. Perhaps the most readily seen signature of infalling gas in low mass 
stars and brown dwarfs is a broad, asymmetric and, usually, strong H$\alpha$ 
emission line, though other diagnostic lines have also been identified 
\citep[e.g.][]{mhc01,jmb03}. 
The presence or absence of accretion signatures, thus, provides a  
valuable probe of inner disk evolution: By comparing the fraction of 
objects with signs of ongoing accretion in star forming regions at different 
ages, it is possible to obtain constraints on gas dissipation timescales. 
Furthermore, the study of accretion also gives us insight into the connection 
between the central star and its circumstellar disk, presumably through a 
magnetic field. 

Here we report on a comprehensive study of disk accretion and related 
phenomena in four nearby associations of young stars: $\eta$ Chamaeleontis 
($\eta$ Cha; $\sim$6 Myr), $\beta$ Pictoris ($\beta$ Pic; $\sim$12 Myr), 
TW Hydrae (TWA; $\sim$8 Myr) and Tucanae-Horologium (Tuc-Hor; $\sim$30 
Myr). Many of these stars have been identified based on their X-ray emission 
and later confirmed as young by the detection of Lithium in optical 
spectra. Their membership in the associations is supported by several 
lines of evidence, including measurements of radial velocities, proper 
motions and distances (when possible). For a recent review of the 
properties of these groups, see \citet{zs04}. By investigating
accretion in those four associations spanning an age range from 6 to 
30\,Myr and combining the results with literature data for younger regions 
\cite[e.g.][]{mjb05}, we are able to determine the fraction of accretors as 
a function of age. The high-resolution, high signal-to-noise, multi-epoch 
spectra that we have collected allow us to distinguish accretors from 
non-accretors reliably, using line shapes of H$\alpha$ and other 
diagnostics. Since chromospheric activity influences the same emission lines
as accretion, ambiguity is unavoidable for a few objects, although in most
cases this problem can be mitigated by combining information from different 
lines and analysing line profiles. In addition, we can identify spectroscopic 
binaries whose complex line profiles may mimic accretion. In addition, by 
deriving projected rotational velocities ($v\sin i$) for our targets, we will 
explore for the first time whether there is any evidence for `disk locking' 
-- i.e., for accretors being preferentially slow rotators -- at these older ages. 
We are also able to investigate time variability of emission lines 
in a number of objects. 

\section{Observations and spectral analysis}
\label{obs}

We have obtained a total of $\sim$650 high-resolution spectra of 
100 likely members of four nearby associations -- $\eta$\,Cha, 
TWA, $\beta$\,Pic and Tuc-Hor -- with the echelle 
spectrograph MIKE at the Magellan Clay 6.5-m telescope on Las Campanas, 
Chile. The data were collected during four observing runs between 2004 
December and 2005 July. The summary observing log is given in
Table~\ref{obsrec}. 

MIKE is a double echelle slit spectrograph, consisting of blue and red arms. 
For this accretion study, we concentrated on the red spectra, with coverage 
from 4900\,\AA\ to 9300\,\AA. With no binning and using the 0\farcs35 slit, our 
spectra have a resolution of $R\sim 60\,000$. The pixel scale was 0\farcs14\,pix$^{-1}$ 
in the spatial direction, and about 24\,m\AA\,pix$^{-1}$ at 6500\,\AA\ in the spectral
direction. One peculiarity of MIKE is that it produces slanted echelle spectra, that is,
the spatial direction of the projected slit is not aligned with the CCD columns.
Moreover, this tilt is wavelength dependent. We therefore developed a customized
software package in ESO-MIDAS to take the slant into account. The details of
the reduction software and procedures used will be described in a forthcoming paper.

Integration times were chosen so that the signal to noise ratio (S/N) $\ga 80$ 
per spectral resolution element at 6500\,\AA, except for the brightest stars
where this would have implied an exposure shorter than 120\,s. In those cases, 
we used the longest exposure time shorter than 120\,s that did not saturate the
detector, giving S/N$\sim$80--500, depending on seeing.

In this first of a series of papers, we focus on accretion-related emission lines 
in these spectra. In particular, we have measured the equivalent width (EW) and
the full-width at 10\% of the peak (see Sect. 3) for the H$\alpha$ emission
line. H$\alpha$ is perhaps the most widely used accretion diagnostic, 
mainly because it is highly sensitive and detectable even in weak accretors. 
However, the H$\alpha$ feature is also affected by chromospheric flares and 
stellar winds, leading to complex and sometimes ambiguous behaviour. 
For the linewidth measurements, the continuum level was determined by a linear fit
to the continuum on both sides of the H$\alpha$ line. By measuring the EW in several 
continuum regions without strong features, we estimate that the uncertainty
of our EW values is on average 0.2\,\AA. For the 10\% widths, the errors
are $\sim 5\,$km\,s$^{-1}$. We also looked for other diagnostics, such as He\,I 
(5876 and 6678\,\AA), O\,I (7773 and 8446\,\AA) and the Ca\,II triplet (8498, 
8542 and 8662\,\AA). We measured the projected rotational velocity 
($v\sin i$) of each of our targets by $\chi^2$ fitting with a ``spun-up'' template 
of a slowly rotating standard. More details on the $v\sin i$ measurements
will be provided in a forthcoming paper.

\section{Accretion signatures} 
\label{accsig}

The strength of the H$\alpha$ line has long been used to distinguish 
classical T Tauri stars or accretors (EW $>$ 10\,\AA) from weak-line T 
Tauri stars or non-accretors (EW $<$ 10\,\AA). It is believed, 
particularly in the context of the magnetospheric accretion scenario, 
that in accretors H$\alpha$ is produced in the gas falling in from the 
disk inner edge on to the star. In 
non-accretors, H$\alpha$ emission originates only from chromospheric 
activity, and thus is generally weaker. Furthermore, the H$\alpha$ 
profiles of accretors tend to be much broader, due to the high 
velocity of the accreting gas as well as Stark broadening, and 
asymmetric, as a result of inclination effects and/or absorption 
by a wind component. 

Since the measured EW of H$\alpha$, and thus the threshold for 
classifying an object as an ``accretor'', depends on the spectral type, 
\citet{wb03} proposed the full width of the line at 10\% of
the peak (hereafter, 10\% width) as a more robust diagnostic. Based 
on the presence or absence of veiling in their stellar spectra, they 
proposed that 10\% width $>$ 270 km\,s$^{-1}$ indicates accretion. 
Using physical reasoning as well as empirical findings, \citet{jmb03}
adopted 200 km\,s$^{-1}$ as a more reasonable 
accretion cutoff for the very low mass regime (i.e., brown dwarfs), 
but cautioned that it should be used in combination with additional 
diagnostics whenever possible. Intriguingly, \citet{ntm04} have 
shown that the H$\alpha$ 10\% width correlates very well with the mass 
accretion rate derived through other means. Thus, the 10\% width appears 
to be not only a good qualitative indicator of accretion but also gives 
a quantitative estimate of the infall rate. Here we use both the EW and 
the 10\% width as diagnostics, because accretion should affect both values. 
On the other hand, if the line profile is broadened due to binarity or fast 
rotation, we expect high 10\% width but the EW should be comparable to those 
of non-accretors. We also investigate the presence or absence of certain other 
emission lines -- in particular O\,I (8446\,\AA), He\,I (6678\,\AA) and Ca\,II 
(8662\,\AA) -- that are often associated with accretion as well 
\citep[see discussion and references in][]{mjb05}. Particularly the He\,I 
(6678\,\AA) emission feature appears to be a good indicator of ongoing accretion: 
As shown by \citet{grh02}, chromospherically active stars often show this
line in emission, but at a very weak level with EW below 0.25\,\AA. For 
our spectra, this is in the range of our detection limit and thus not significant. 
In a few cases this line has been reported to be in emission during 
chromospheric flare events \citep{msc98,ma01} with EW of a few\,\AA; thus
attributing non-persistent emission in this line to accretion might be
problematic. However, if we clearly detect He\,I (6678\,\AA) in emission in
all spectra for a given star, the object is very likely accreting.

We have measured the H$\alpha$ EW and 10\% width in all the spectra in 
hand for all late-type stars in our sample (see Sect. \ref{obs}). For objects 
where we see H$\alpha$ in emission, the average values and their standard deviations 
are reported in Tables \ref{res1}-\ref{res4}; the $\sigma_\mathrm{EW}$ gives an 
indication of the line's variability. We note that these values are in many 
cases clearly higher than our measurement uncertainties (see Sect. \ref{obs}), 
demonstrating that many stars in our sample show emission line variability (see Sect. 
\ref{variab}). The tables also indicate the presence of the He\,I (6678\,\AA) 
emission feature, which is an additional diagnostic for ongoing accretion. In 
the four panels of Fig. \ref{fig1}, we plot the H$\alpha$ EW 
vs. 10\% width for members of each of the four stellar associations. This plot 
is a good way to identify likely accretors in each group, using the criteria 
discussed above. We show H$\alpha$ profiles for selected objects in Fig. 
\ref{hats1} and \ref{hats2}.

Among $\eta$ Cha cluster members, one star -- ECHA J0843.3-7905 or $\eta$ 
Cha 13 -- shows strong signs of accretion, with large H$\alpha$ 
equivalent and 10\% widths, comparable to those seen in classical T 
Tauri stars, and a host of other broad emission lines. $\eta$ Cha 9 also 
makes the cut as an accretor, based on both its EW and 10\% width as well 
as the presence of He\,I (6678\,\AA). Two other members -- $\eta$ Cha 7 and 11 -- 
exhibit broad lines wings, with 10\% widths above 300\,km\,s$^{-1}$, but 
their emission is relatively weak (Fig. \ref{hats1}). \citet{llf04} found 
$\eta$ Cha 7 to be a double-line spectroscopic binary. Our H$\alpha$ profiles 
of it could well result from the blending of two emission lines from the two 
binary components (Fig. \ref{hats1}). Besides, it does not show any other 
accretion-related emission. Thus, we conclude that $\eta$ Cha 7 is not 
an accretor. The H$\alpha$ profile of $\eta$ Cha 11 also appears to consist 
of more than one component (see Fig. \ref{hats1} and Sect. \ref{variab}), but 
this target does not show any other signs of binarity \citep{zs04}. Whether it 
is a binary or not, the broad red absorption feature in its H$\alpha$ 
profile, seen at several epochs, is hard to explain without invoking 
high-velocity infalling material; therefore, we classify $\eta$ Cha 11 
as an accretor. The star $\eta$ Cha 5 is at the boundary between accretors
and non-accretors. Of our five spectra of $\eta$ Cha 5, only one shows 
broad H$\alpha$ (Fig. \ref{hats1}). The average 10\% width is $\sim$200 
kms$^{-1}$, i.e. well below the \citet{wb03} threshold. Furthermore, none 
of its spectra shows other definitive signs of accretion such as 
He\,I (6678\,\AA) emission. Given these findings, we conclude that 
it is not accreting during our observations. On the other hand, 
\citet{llm04} detect broad H$\alpha$ emission with more than
300\,kms$^{-1}$ 10\% width in a single spectrum and conclude that $\eta$ Cha 5
is accreting at a significant level. Thus, the status of this object is questionable;
it appears to show some sporadic accretion, indicating that accretion
variability can be quite substantial on long timescales. Since it is 
non-accreting in our multi-epoch spectra, we do not count it as an accretor
in the remainder of this paper.


Both ground-based and Spitzer mid-infrared data are available for members of the 
$\eta$ Cha group. The L'-band (3.8\,$\mu$m) survey by \citet{hja05} found infrared 
excess for only two stars, $\eta$ Cha 11 and 13, both of which we classify as 
accretors above. The Spitzer IRAC photometry by \citet{mhl05} indicates 4-10$\mu$m 
excess emission for four stars in our sample, $\eta$ Cha 5, 9, 11, and 13 (called
15 in their paper), where $\eta$ Cha 5 and 9 show excess only at wavelengths 
$>4\,\mu$m. The IRAC colour excess for $\eta$ Cha 5 confirms that it still might
be able to accrete sporadically, as seen by \citet{llm04}, but not in our spectra.
It is intriguing that there is no L'-band, but IRAC colour excess plus evidence 
for, at least sporadic, accretion in $\eta$ Cha 5 and 9. This could imply
substantial grain growth and/or partial clearing of the inner disk, while they are
still accreting. Thus, these objects might have inner disks depleted of dust, but
with substantial amounts of gas. This may also be the case with the accreting
star TW Hydrae (see below). 

In the TW Hydrae association, two members -- TW Hya and Hen 3-600A (TWA 3A) 
-- stand out as accretors based on their H$\alpha$ emission as well as the 
presence of other emission lines such as He\,I (6678\,\AA). Both had been 
identified before, though not in a systematic survey \citep{m00,mjb03}. 
Based on mid-infrared photometry of TWA members known at the time, 
\citet{jhf99b} reported excesses consistent with inner disks 
for these two accretors \citep[as well as the HD 98800 quadruple system and 
the A star HR 4796A; also see][]{j98}. Recent Spitzer photometry indicates that
for both stars the SED does not turn over even by 160$\,\mu \mathrm{m}$, again consistent with
their status as stars with active accretion disks \citep{lsw05}. While TW Hya is 
accreting, shows significant mid-infrared excess, and has a face-on disk resolved 
in scattered light \citep[e.g.][]{k00} and in the millimeter \citep{wbw03}, it 
appears to lack near-infrared excess. Based on model fits to the spectral energy 
distribution (SED) and the millimeter image, there is evidence for partial 
clearing of the innermost few AU of its disk \citep{c02}. In the case of the 
Hen 3-600 binary system with roughly equal mass companions (separation 1\farcs4 or 
$\sim$70 AU), it is interesting that only the primary shows signs of 
accretion as well as mid-infrared excess \citep{jhf99a}. 

Two more objects show emission line characteristics typical for accretors
in at least a few spectra:
TWA 10 makes the cut as an accretor based on its EW, but the 10\% width is
at the borderline. A closer examination of the line profiles shows a burst-like 
event in our time series (see Sect. \ref{variab}). During 
this event, it has increased linewidth in H$\alpha$, He\,I (6678\,\AA) emission 
as well as a strongly asymmetric profile, which is 
often seen in accreting T Tauri stars. Another interesting case is TWA 5A: This 
object has broad H$\alpha$ emission,
but again the values are influenced by a burst event in our time series. 
During this burst, both EW and 10\% width clearly exceed the adopted border
between accreting and non-accreting objects. In the remaining spectra,
the 10\% width is still very high, but the EW is around or close to 10\,\AA.
Again, He\,I (6678\,\AA) is detected, but only during the burst event. We
examine the H$\alpha$ variability of TWA 5A in detail in Sect. \ref{variab}.
TWA 5A has been suspected to be an accretor previously, based on its broad and
variable H$\alpha$ emission \citep{mjb03}.

Although both TWA 10 and TWA 5A show emission line spectra resembling
features seen in accreting stars, the recent analysis of their infrared SEDs
obtained from Spitzer casts doubts on their spectroscopic classification
as accretors \citep{lsw05}: For both objects, the fluxes (or upper limits) at 
24, 70, and 160$\,\mu \mathrm{m}$ are consistent with pure photospheric emission, without
any evidence for excess emission due to disks. This clearly separates them from 
the accretors TW Hya and Hen 3-600, whose infrared luminosities are enhanced by several 
orders of magnitudes. Given the non-detection of a disk at mid- and far-infrared wavelengths 
with the high sensitivity of Spitzer, ongoing accretion is unlikely
for these two objects. Thus we classify both of them as non-accretors.
The striking, but not persistent emission line spectrum is in both cases probably 
due to magnetic activity: As discussed above, chromospheric flares can possibly
lead to He\,I (6678\,\AA) as well as broad H$\alpha$ emission. Although it 
is still doubtful how a flare can also account for the profile variations
seen in TWA 10 and TWA 5A, chromospheric activity appears to be the best explanation 
for the observed behaviour. The extremely broad H$\alpha$ line of
TWA 5A is additionally affected by its fast rotation (see Sect. \ref{rotacc}) and
binarity: The object is a known roughly equal-mass AO binary \citep{bjn03}, and there 
are indications for one of the components being a spectroscopic binary itself \citep{t03}. 
These two cases demonstrate that caution has to be applied when interpreting emission 
line spectra of young stars. Multi-epoch, multi-wavelength data have to be used for
a reliable characterization of T Tauri stars.

Five other objects -- TWA 6, 14, 17, 19B and 20 -- have large 10\% widths, but the 
emission is rather weak. Our multi-epoch spectra show evidence of binarity 
in all five cases (e.g., double-peaked Li\,I 6708\,\AA), suggesting that 
line broadening is the result of the blending of two components rather than 
high-velocity infalling gas. Perhaps not surprisingly, the same five, along 
with TWA 5A, have the largest measured $v\sin i$ in the group (see Tables 
\ref{res1}-\ref{res4}), probably due to blending of absorption lines of 
the two components in the spectra. Furthermore, complementary accretion
indicators like He\,I emission are absent (or seen in absorption) in the spectra 
of these five objects. In summary, we conclude that TW Hya and Hen 3-600A
are the only accreting stars in our TWA sample. Additionally, one of the four 
young brown dwarfs that are likely members of the TWA, 2MASSW J1207334-393254, 
is also known to be accreting \citep{g02,mjb03}. We have analyzed its variable 
accretion signatures in two recent papers \citep{sjb05,sj06}.


Of the $\beta$ Pic moving group members in our sample, none shows 
definitive evidence of on-going accretion. CD-64 1208 does have a large 
10\% width, but it is clearly a spectroscopic binary, with line-blends  
accounting for the broad wings; its Li\,I absorption line is double-peaked. 
The H$\alpha$ EW is very small and it does not exhibit any other emission 
lines associated with accretion. We note that AU Mic, which has a 
resolved dust disk \citep{k04}, does not appear to be accreting. 

Among Tuc-Hor moving group members in our sample, two -- CD-53544 and 
HIP 2729 -- have large H$\alpha$ 10\% widths. We do not find evidence 
for binarity in either case, but both are fast rotators, with $v\sin i$ 
$\sim$80 and $\sim$130 km\,s$^{-1}$, respectively. Neither exhibits 
other accretion-related emission lines. The photospheric lines are 
clearly broadened by rotation, and the same could account for the broad 
H$\alpha$. Thus, we do not find any accretors in the Tuc-Hor group either. 
For this group, \citet{mmh04} carried out a mid-infrared survey, but
did not find excess emission for any star, indicating that inner 
circumstellar disks are optically thin and undetectable at the age of 
Tuc-Hor.

\section{Frequency and lifetime of accreting disks}
\label{life}

Based on our comprehensive, multi-epoch spectroscopic survey, supplemented 
with previously published results, we find that three of the 11 late-type 
stars in $\eta$ Cha ($27\pm_{14}^{19}$\%) and two of the 32 targets in the 
TWA ($6\pm_4^7$\%) show evidence of on-going accretion. None of the 21 $\beta$ 
Pic and 36 Tuc-Hor targets in our survey appears to be accreting. This
corresponds to upper limits on the accretor frequency of 13\% for $\beta$ Pic 
and 8\% for Tuc-Hor (95\% confidence). 
As shown in Fig. \ref{freq}, these accretor fractions are well below those found 
for younger ($\sim$1--5 Myr) star-forming regions, where the values 
are generally higher than 30\% \citep{mjb05}. Except in the case of $\eta$ Cha 
(where we are necessarily limited by the small number of known members), 
these lower accretion disk frequencies at older ages can no longer be 
dismissed as being based on very small samples. Furthermore, the fact that 
we have collected multi-epoch spectra for our targets reduces (but does not 
eliminate) the likelihood that we have missed some non-steady accretors in 
this sample. 

To probe the evolution of accretion in these young groups, a critical
assessment of their ages is important. For $\eta$ Cha, we adopt the most
recent age estimate of $6\pm_1^2$\,Myr, based on comparison of the observed HR 
diagram with evolutionary tracks \citep{ls04}. This value is consistent
with previous estimates. \citet{w99} discuss different age measurements
for the TW Hydrae group and conclude that the most probable age is $\sim 8$\,Myr.
Recently \citet{lc05} found that the TW Hydrae group has two spatially
distinct subgroups, whose ages differ by a factor of two. Based on rotation
periods and HR diagrams, they argue that the objects TWA 14--19 are likely
by a factor of two older than the objects TW Hydrae and TWA 2--13. 
(TWA 20--25 are not included in their study.) From the Li EW, probably the most
reliable observational signature to assess (relative) ages of young stars, there
is, however, no evidence for two subgroups with significantly different ages in TW 
Hydrae \citep{zs04}. At least the M-type objects TWA 14, 15, and 18 should have
depleted their Lithium if they were much older than 10\,Myr, which
is not seen in the data \citep[see][their Fig. 3]{zs04}. Additionally, the 
comparison between the rotation periods of the potential subgroups is clearly 
hampered by small-number statistics. Thus, we decided to treat TW Hydrae as a 
uniform group with an age of $\sim 8$\,Myr.

For the $\beta$ Pic moving group, both kinematic and evolutionary track
diagnostics give an age of $\sim 12$\,Myr \citep{zs04}. For these three 
groups, the age uncertainty is probably a few Myr. However,
using the Li abundance as an age indicator, \citet[][their Fig. 3]{zs04} have
convincingly shown that $\beta$ Pic is clearly older than TW Hydrae, and 
that TW Hydrae is older than $\eta$ Cha. Finally, the most
probable age for Tuc-Hor is $\sim 30$\,Myr \citep{zs04,t00}.

By combining these age estimates with our accretor frequencies, we can set
strong constraints on the lifetime of accreting disks: disk accretion appears 
to cease or dip below measurable rates by about 10 Myr for most late-type stars. 
Longer-lived accretors must be rare, given that we did not find any among 57 targets 
in $\beta$ Pic and Tuc-Hor groups. At $\sim$12 Myr, the estimated age of the 
$\beta$ Pic group, the accretor fraction drops below 13\% (95\% confidence limit); 
at $\sim$30 Myr, the age of Tuc-Hor, it is $<8$\%. These numbers are in full
agreement with constraints from previous studies, which find that accretor
frequency and accretion rates drop significantly between 4 and 10\,Myr
\citep[e.g.][]{m00,shh05}. This also provides an indirect constraint on the timescale 
for gas dissipation in the inner disks and, in turn, on the timescale for gas 
giant planet formation. Interestingly, the lifetime we derive for gas accretion is 
roughly consistent with that found for dust in inner disks through previous near- 
and mid-infrared studies \citep[e.g.][]{hll01,jhf99b}.

\section{Accretion-rotation connection?}
\label{rotacc}

Over the years, a number of studies have investigated the connection between 
the presence of disks and stellar rotation. In the magnetospheric accretion scenario, 
gas from the inner disk is thought to be channeled by the stellar magnetic field. 
In that case, one would expect the field lines to connect the star to the disk,
which may prevent the star from spinning up (or at least reduce the spin up) as 
it contracts during the pre-main sequence phase. This process is often referred 
to as ``disk locking'' \citep{c90,k91,s94}. In other words, we might expect 
accreting stars to be preferentially slow rotators compared to their peers.
Indeed, initial studies by Edwards et al. (1993) found a correlation between rotation 
period and near-infrared colour excess, interpreted as a connection between
star and disk. Although this result has later been confirmed, e.g. by \citet{h02}, 
other groups fail to detect such a correlation \citep[e.g.][]{s99}.  

One problem with many studies of the disk-rotation connection is the use 
of near-infrared excess as the disk diagnostic. Due to a variety of effects,
e.g. dust settling and inner disk clearing, near-infrared photometry is not a
very robust disk indicator. Mid-infrared data are much more reliable for 
this purpose, and are now available for most of our targets \citep{jhf99b,lsw05,mhl05}.
Studies with the aim of investigating the connection between mid-infrared
emission and rotation are underway \citep[e.g.][]{rsm05}. However, both near-infrared
and mid-infrared signatures probe the {\it presence} of a dusty disk, and
not the {\it coupling} between star and disk, which is required in the disk
locking scenario. As shown here and elsewhere \citep[e.g.][]{j01}, the existence 
of a disk does not necessarily imply a star-disk connection, especially at slightly 
older ages. Therefore, evidence of on-going accretion -- which signals a direct link 
between the inner disk and the central star -- is a much more sensible diagnostic 
to correlate with stellar rotation rates. 

At present, there has been to the best of our knowledge no study of a possible 
accretion-rotation connection at ages of $\sim$6--30 Myr. This is a difficult task, 
because only few accretors are present at these ages, hampering a reliable statistical
comparison of the rotational properties of non-accretors and accretors. Nevertheless,
we decided to probe disk-locking at 6-8\,Myr using our comprehensive dataset for $\eta$ 
Cha and TWA. From previous investigations using near-infrared colour excess as 
disk diagnostic, we expect to see three groups of objects, if disk-locking is still
at work in our objects \citep[see review by][]{hem06}: a) Slowly rotating stars with 
disk, b) slowly rotating stars without disk, c) fast rotating stars without disk. In 
some sense, this corresponds to an evolutionary sequence: Stars rotate slowly, as long
as they are coupled to their disks. After losing the disk, they still rotate slowly
for a certain time, because it takes some time to spin them up. Eventually, they will
become fast rotating objects without disk, after rotational acceleration due to 
pre-main sequence contraction. For this simple picture, the critical test is if 
there are {\it fast rotators with disks} -- those objects should not exist in a disk-locking 
scenario, and the most recent investigations of this issue confirm this expectation for 
1\,Myr old objects based on near- and mid-infrared photometry \citep{h02,rsm05}. Here 
we explore if the same behaviour still holds at 6-8\,Myr.

Fig. \ref{vsinivsten} shows the H$\alpha$ 10\% width (an accretion diagnostic) vs. the 
projected rotational velocity ($v\sin~i$) for $\eta$ Cha and TW Hydrae stars. Since faster 
rotation would contribute to line broadening, it is not surprising that there appears 
to be an overall positive correlation between $v\sin~i$ and 10\% width.
In $\eta$ Cha, there is no obvious difference between accretors and non-accretors: 
all stars are relatively slow rotators with $v \sin i \sim <$ 20\,km\,s$^{-1}$ (with one
exception, $\eta$ Cha 9, but here the 10\% width is clearly influenced by binarity, 
see Sect. \ref{variab}). In the scenario discussed above, we are missing the fast rotators
without disk. This may simply indicate that the stars have not had enough time to 
spin up. In the somewhat older TW Hya group, the two bona fide accretors -- TW Hya 
and Hen 3-600A (TWA 3A) -- are both slow rotators with $v\sin i <$ 15\,km\,s$^{-1}$. 
All other non-accreting objects cover the whole range in rotational velocities with 
$v\sin i$ up to 50\,km\,s$^{-1}$ (excluding spectroscopic binaries with blended line 
profiles). Thus, the available data for the $\eta$ Cha and TW Hydrae group are consistent 
with the described disk-locking scenario. Since there are only a few accretors in those 
two groups, however, small number statistics are a big concern. 

In $\eta$ Cha and TW Hydrae, rotation periods have been derived for a few objects 
\citep{lc05,lcm01}, including some accretors, through photometric monitoring. In 
both groups, the periods confirm our results from the $v\sin i$ analysis, albeit again
based on very few objects. With one exception, all periods measured in $\eta$ Cha are 
in the range between 1 and 20\,days, and the two accretors with periods ($\eta$ Cha 9 
and 11) are indistinguishable from the non-accretors. In TW Hydrae, \citet{lc05} find
two rotationally (and spatially) distinct groups of stars (see Sect. \ref{life}). 
We do not see this dichotomy in our $v\sin i$ data, particularly not after excluding
the binaries, which have the highest measured $v\sin i$ values (see Sect. \ref{accsig}), 
In any case, the accretor with known period (TW Hydrae itself) belongs to the group  
of slow rotators with a period of 2.8\,d, consistent with our result from the $v\sin i$ 
vs. 10\% width plot.

In summary, we find that all accretors in our sample are slow rotators 
with $v \sin i \lesssim$ 20\,km\,s$^{-1}$, whereas non-accretors show a large spread in
rotational velocities up to $50\,km\,s^{-1}$. This is consistent with a scenario
where rotational braking by coupling between star and disk can operate in low-mass stars 
for up to $\sim$8 Myr. Given the small numbers of accretors at these ages, however, this
result is of low significance. It clearly needs to be checked with larger samples 
if/when they become available.

\section{Emission Line Variability}
\label{variab}

Many previous studies of young stars have provided evidence for significant variability
in the emission lines, particularly H$\alpha$, both in intensity and in profile 
shape \cite[e.g.][]{jb95,ab02,sjb05}. This was one of the reasons to base our
accretor frequency analysis on multi-epoch spectra, to avoid signficant bias
due to variability. In some cases, variability information can be used to
obtain a detailed view on the accretion behaviour of young stars. Here we want 
to use our spectra to assess the H$\alpha$ variability of our targets. 

\begin{figure}[t]
\begin{center}
\includegraphics[width=13.5cm,angle=0]{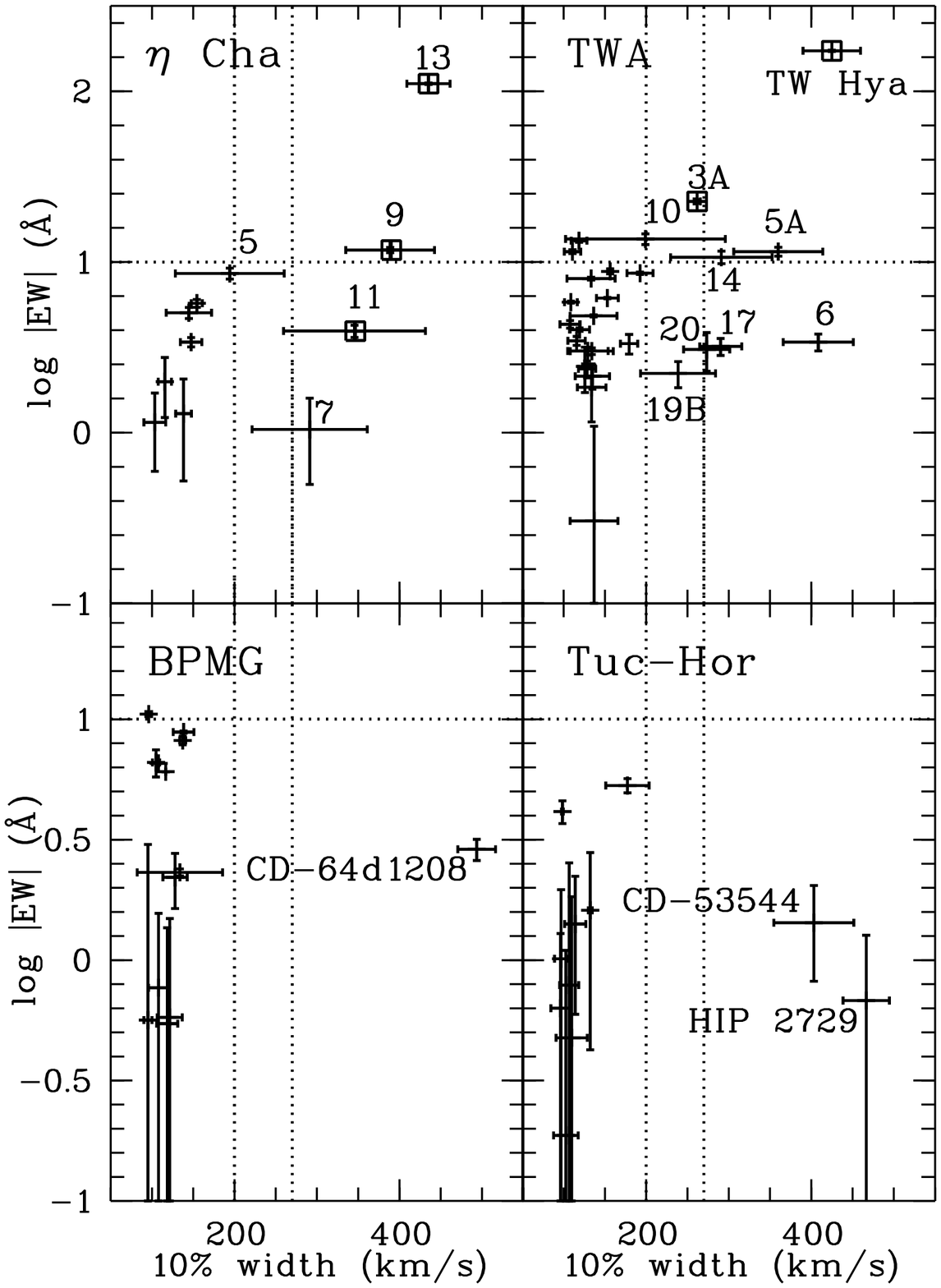}
\caption{H$\alpha$ 10\% width vs. equivalent width for all four groups. The adopted boundaries
between accretors and non-accretors are shown as dotted lines (see discussion in Sec. 
\ref{accsig}). Specific objects discussed in the text are labelled.
Objects classified as accretors in Sec. \ref{accsig} are marked with squares. 
The 'errorbars' do not correspond to the measurement uncertainty, which is 
$\sim 0.2$\,\AA in EW and $\sim 5\,$km\,s$^{-1}$ in 10\% width, but to the scatter 
in our multi-epoch data.
\label{fig1}} 
\end{center}
\end{figure}

\begin{figure}[t]
\begin{center}
\includegraphics[width=12.5cm,angle=-90]{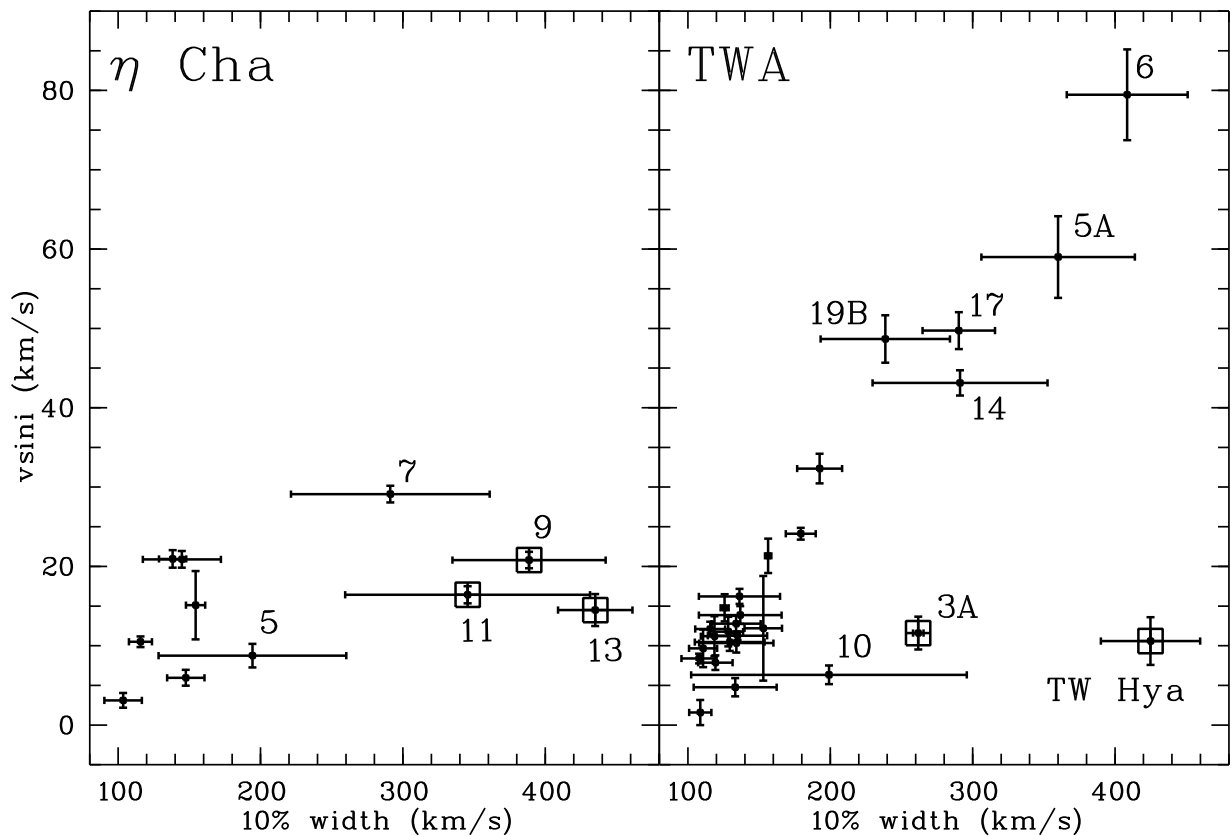}
\caption{Projected rotational velocity $v\sin~i$ vs. H$\alpha$ 10\% width for the groups
$\eta$ Cha and TW Hydrae. Specific objects discussed in the text are labelled.
Objects classified as accretors in Sect. \ref{accsig} are marked with squares. The 'errorbars' 
do not correspond to the measurement uncertainty, but to the scatter in our multi-epoch data 
(see discussion in Sec. \ref{rotacc}).  
\label{vsinivsten}} 
\end{center}
\end{figure}

\begin{figure}[t]
\begin{center}
\includegraphics[width=10cm,angle=-90]{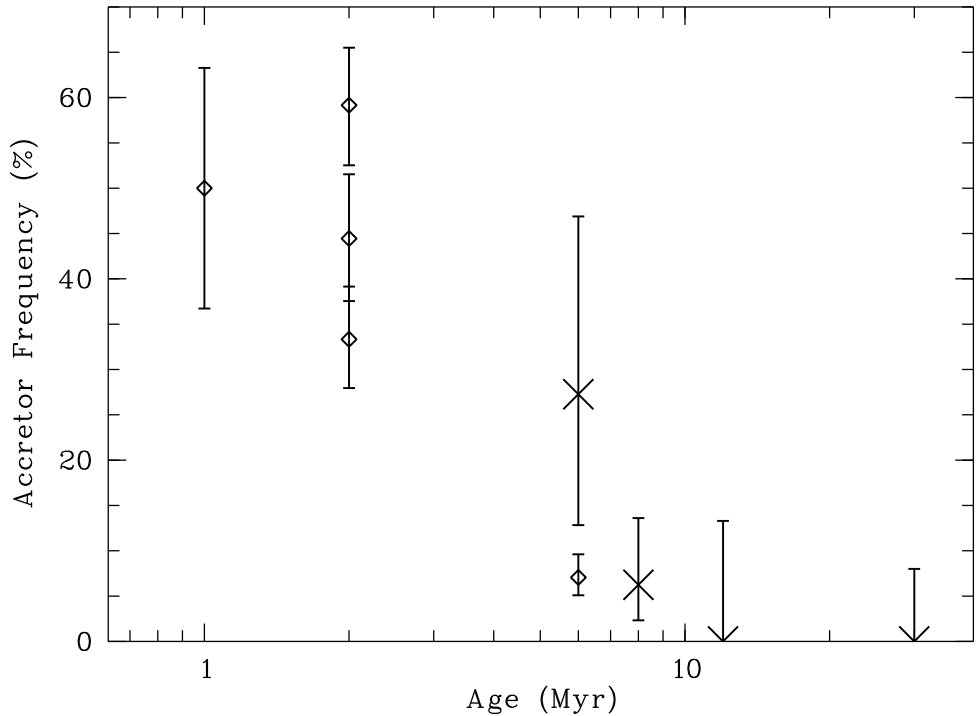}
\caption{Accretor frequency as a function of age. Diamonds mark values from \citet{mjb05}
for K0-M4 stars in different star forming regions, i.e. objects with comparable spectral type,
but lower ages than our targets. Crosses are the datapoints for $\eta$ Cha,
TW Hydrae, $\beta$ Pic, and Tuc-Hor from our study. Errorbars are 1$\sigma$ uncertainties
for detections and 95\% confidence upper limits for non-detections. Age uncertainties
are $\sim 2$\,Myr except for Tuc-Hor, where the error is more likely 5\,Myr. The {\it relative} 
ages for our four target groups are more reliable than absolute values. (see Sect. \ref{life}). 
\label{freq}} 
\end{center}
\end{figure}

\begin{figure}[t]
\begin{center}
\includegraphics[width=7cm,angle=0]{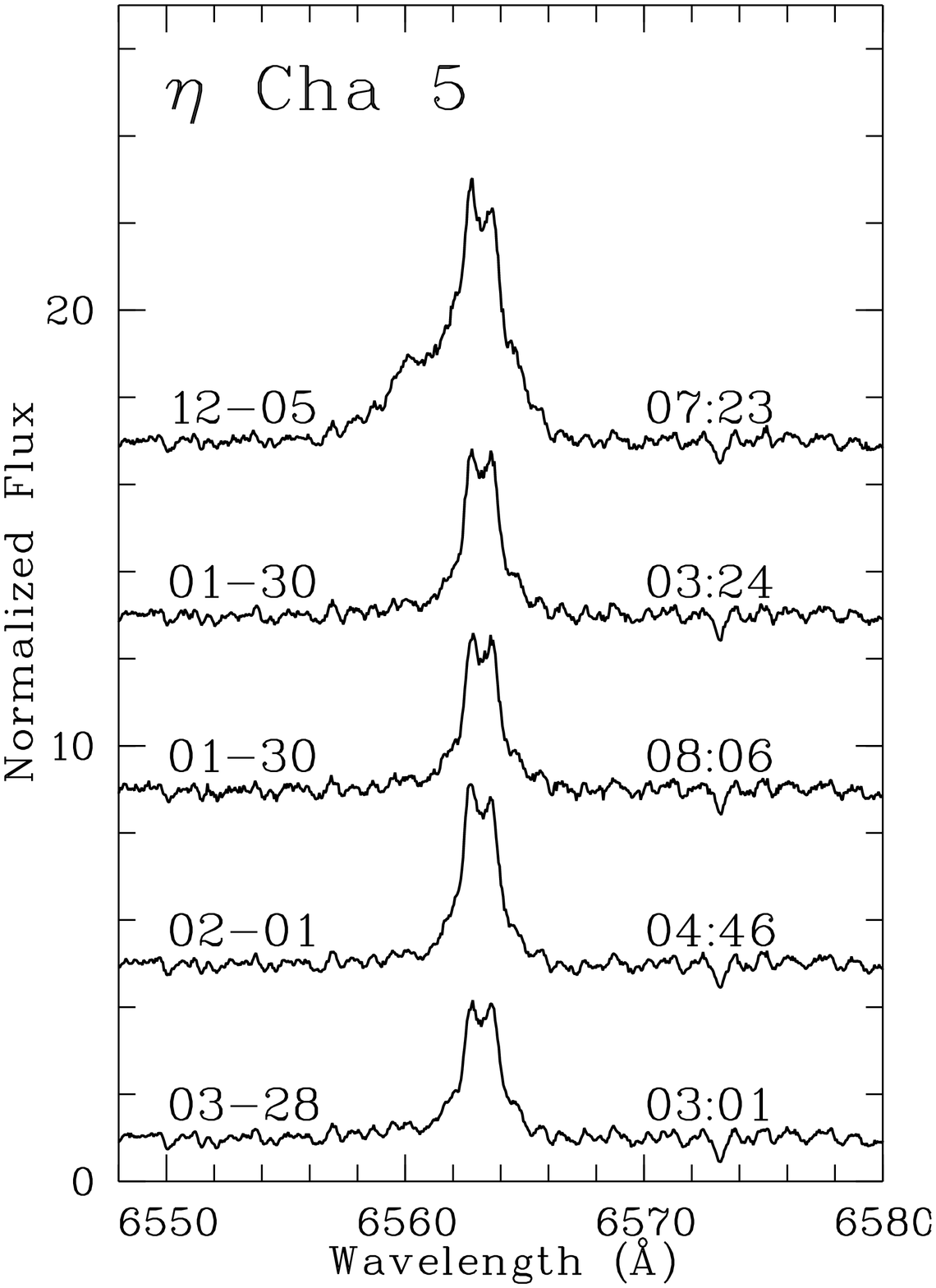}
\includegraphics[width=7cm,angle=0]{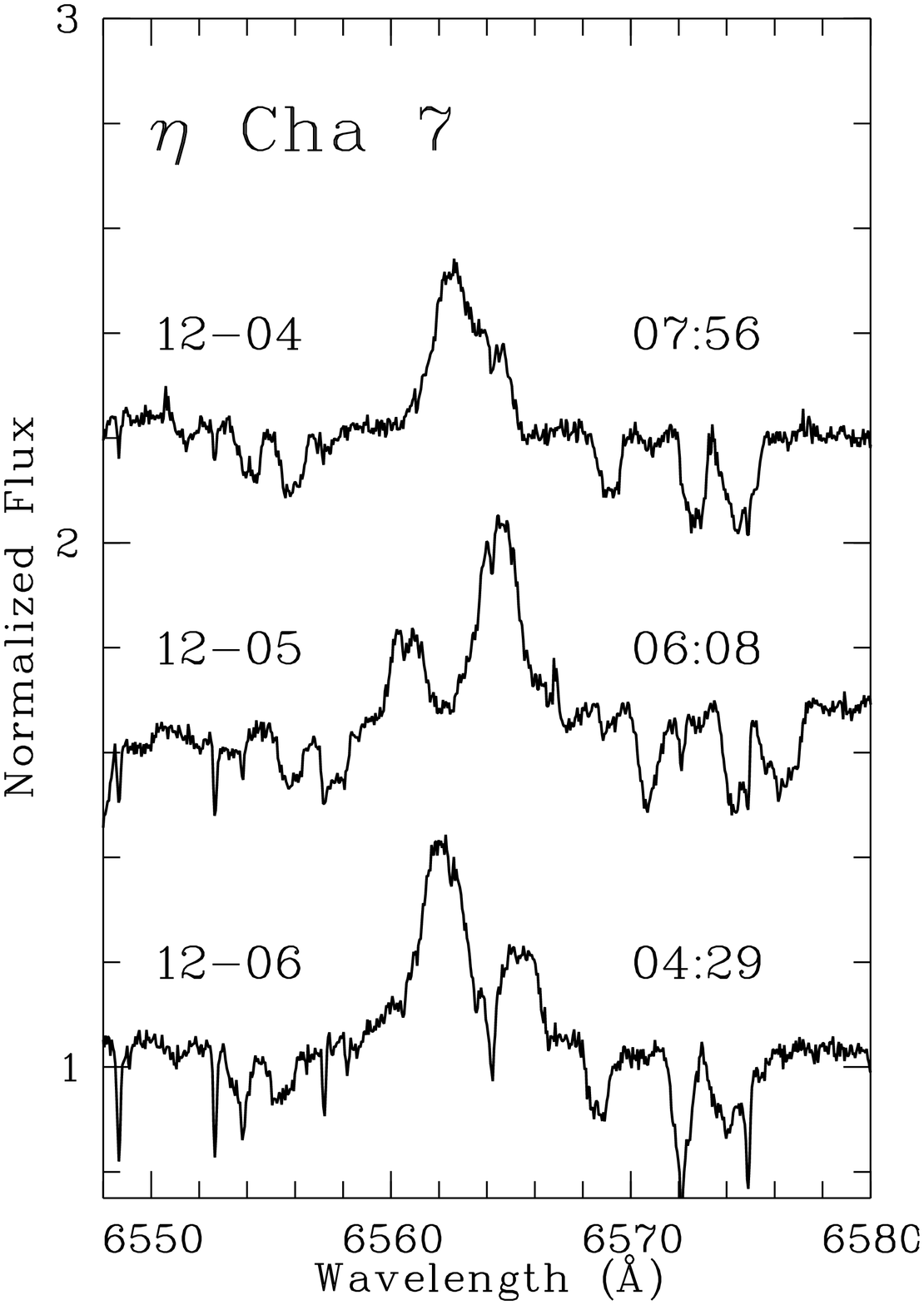}\\
\includegraphics[width=7cm,angle=0]{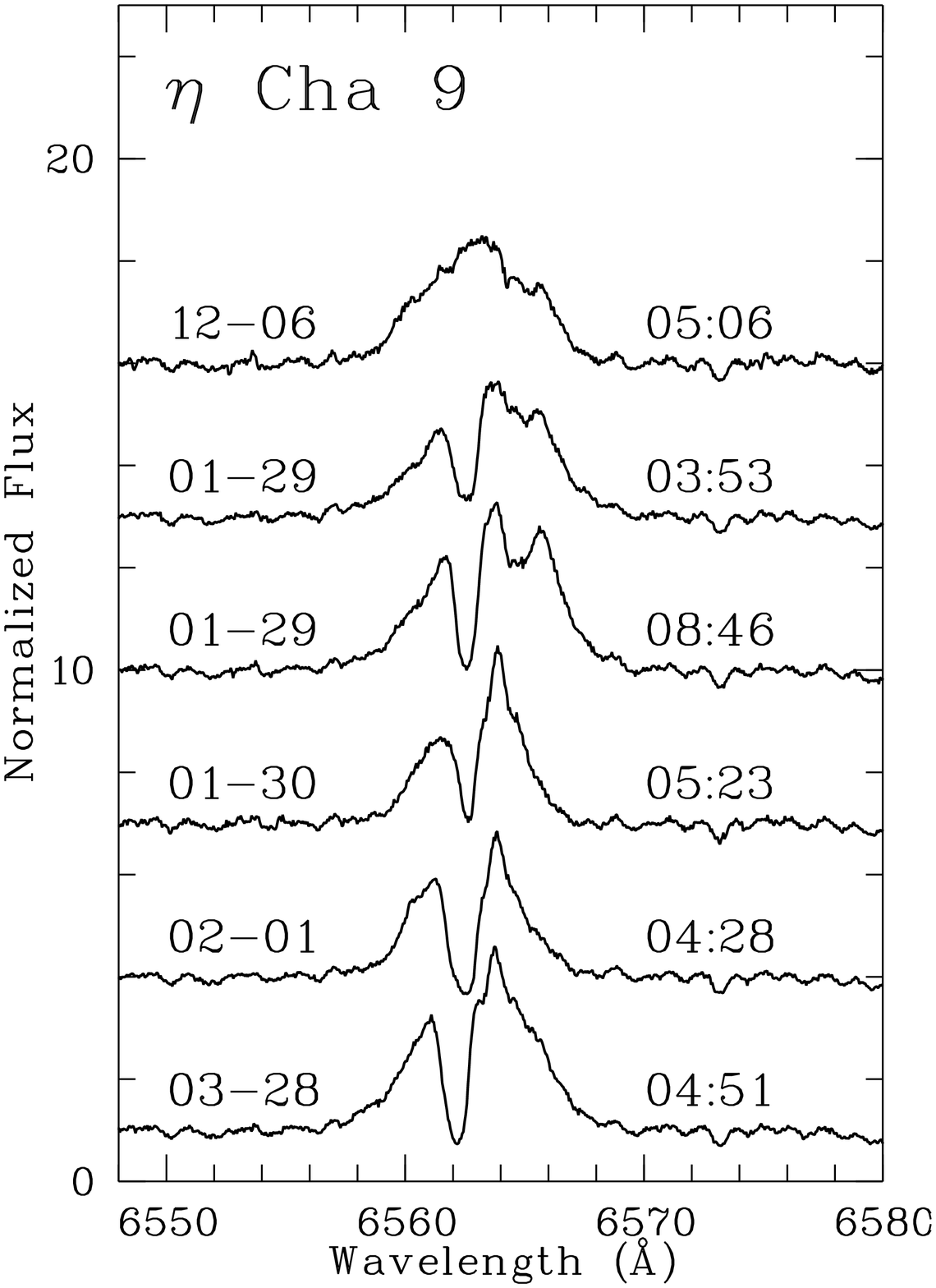}
\includegraphics[width=7cm,angle=0]{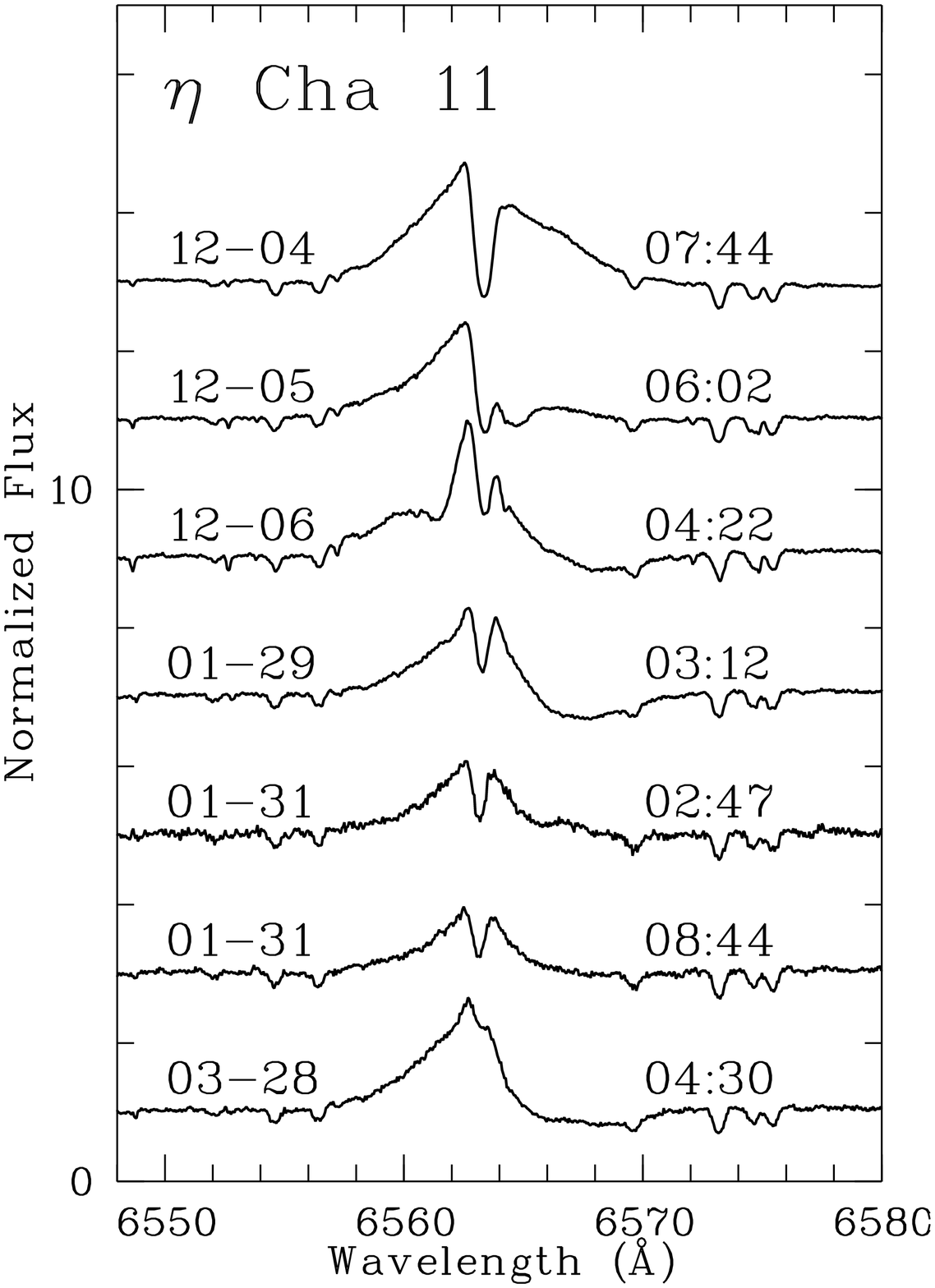}
\vspace{0.5cm}
\caption{H$\alpha$ time series for selected stars in the $\eta$ Cha group in chronological
order (from top to bottom) with UT date and time. Profiles from the different epochs have been shifted
by arbitrary units for clarity. All profiles are normalised to the continuum. 
\label{hats1}} 
\end{center}
\end{figure}

As indicated by the ``errorbars'' in Fig. \ref{fig1}, many of our 
targets show significant variability in the H$\alpha$ line in our multi-epoch data. 
Note that these ``errorbars'' do not show the {\it uncertainty} of our measurements 
of EW and 10\% width, but the {\it scatter} of the individual measurements of our 
multi-epoch data. In most cases, this scatter, given in Tables \ref{res1}--\ref{res4},
is much higher than the formal error. Thus, H$\alpha$ variability is a common phenomenon 
in the objects in our sample, comparable to the results obtained previously for 
younger T Tauri stars \citep[e.g.][]{jb95,ab02}.

\begin{figure}[t]
\begin{center}
\includegraphics[width=7cm,angle=0]{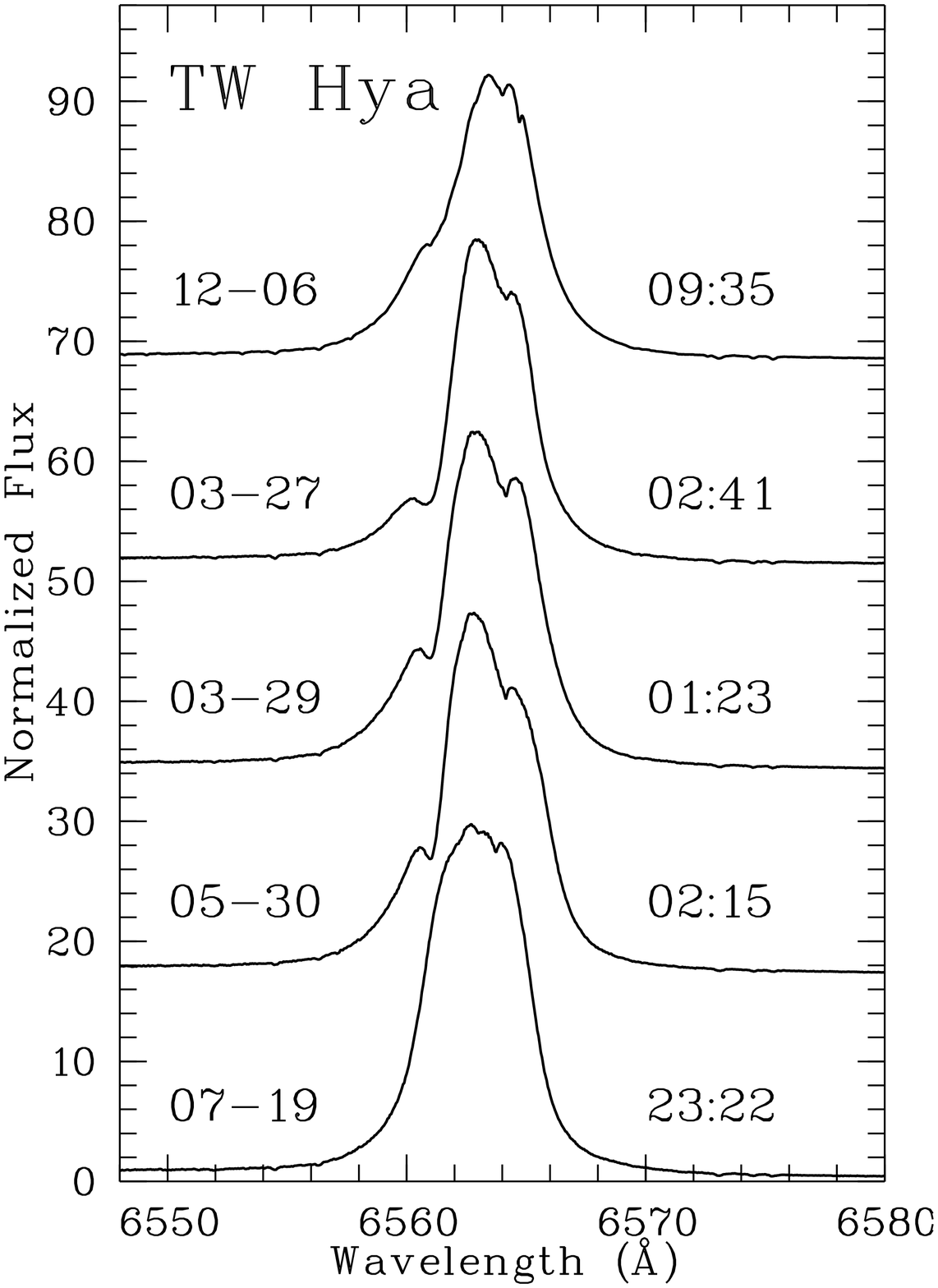}
\includegraphics[width=7cm,angle=0]{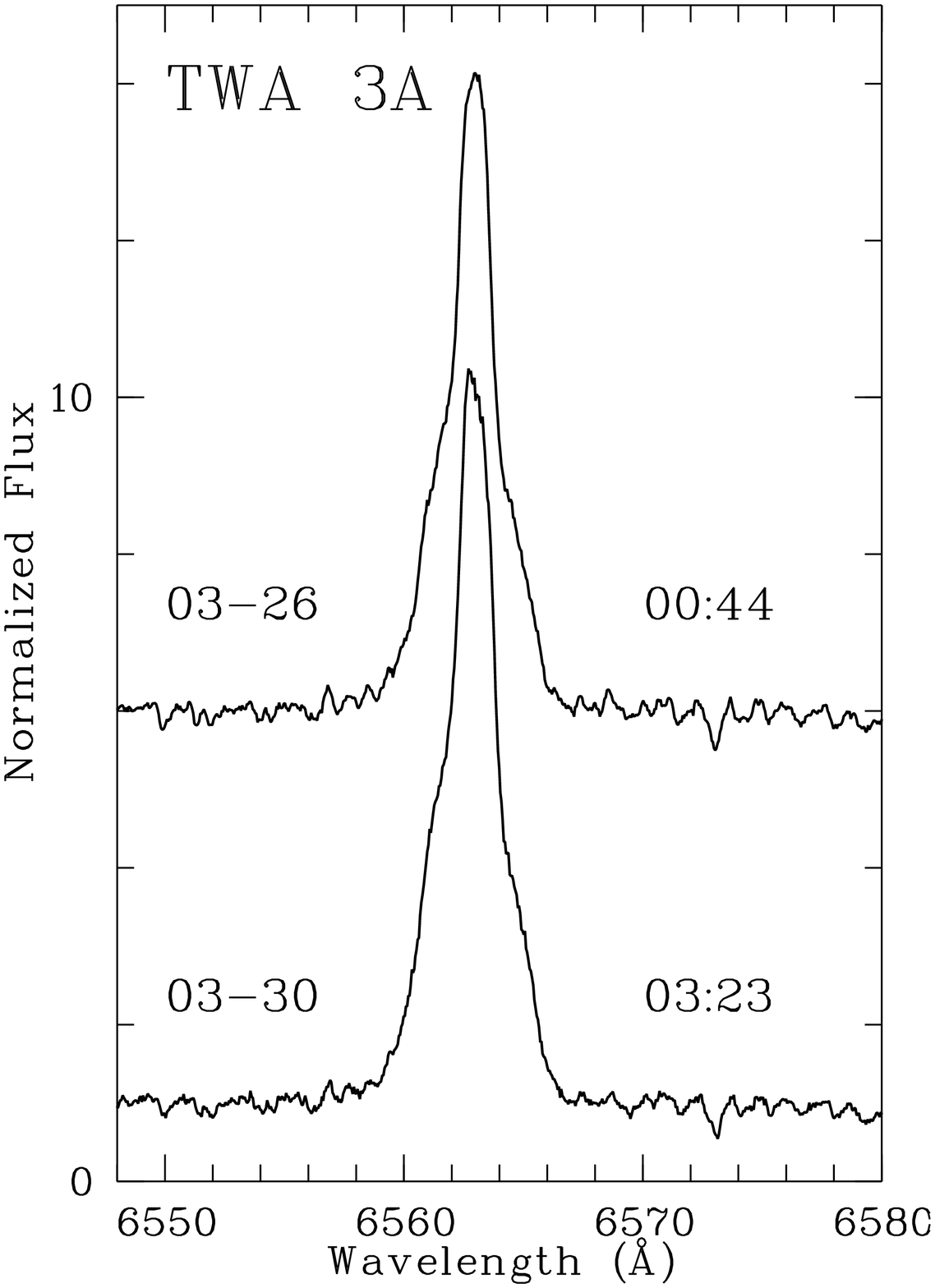}\\
\includegraphics[width=7cm,angle=0]{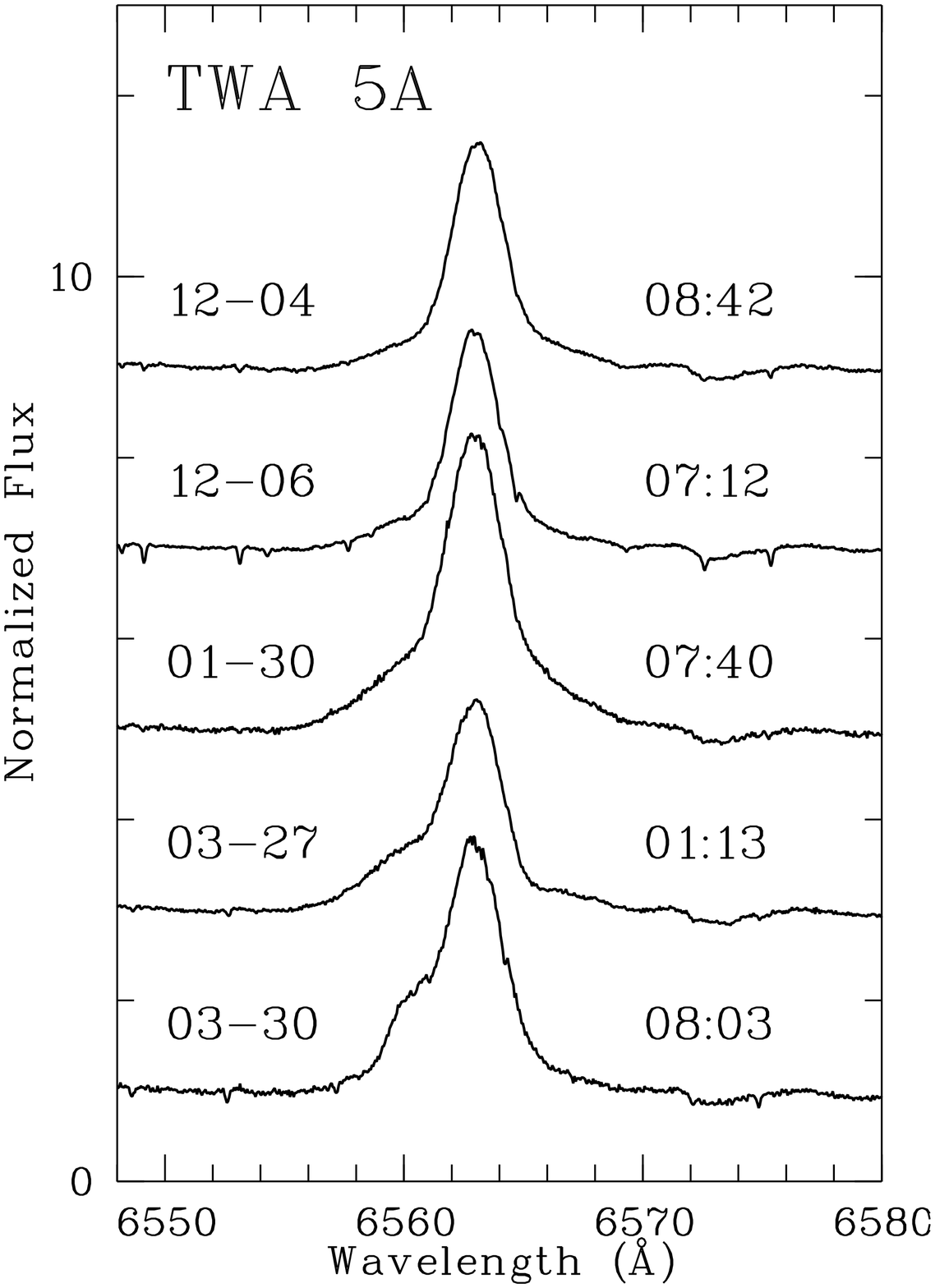}
\includegraphics[width=7cm,angle=0]{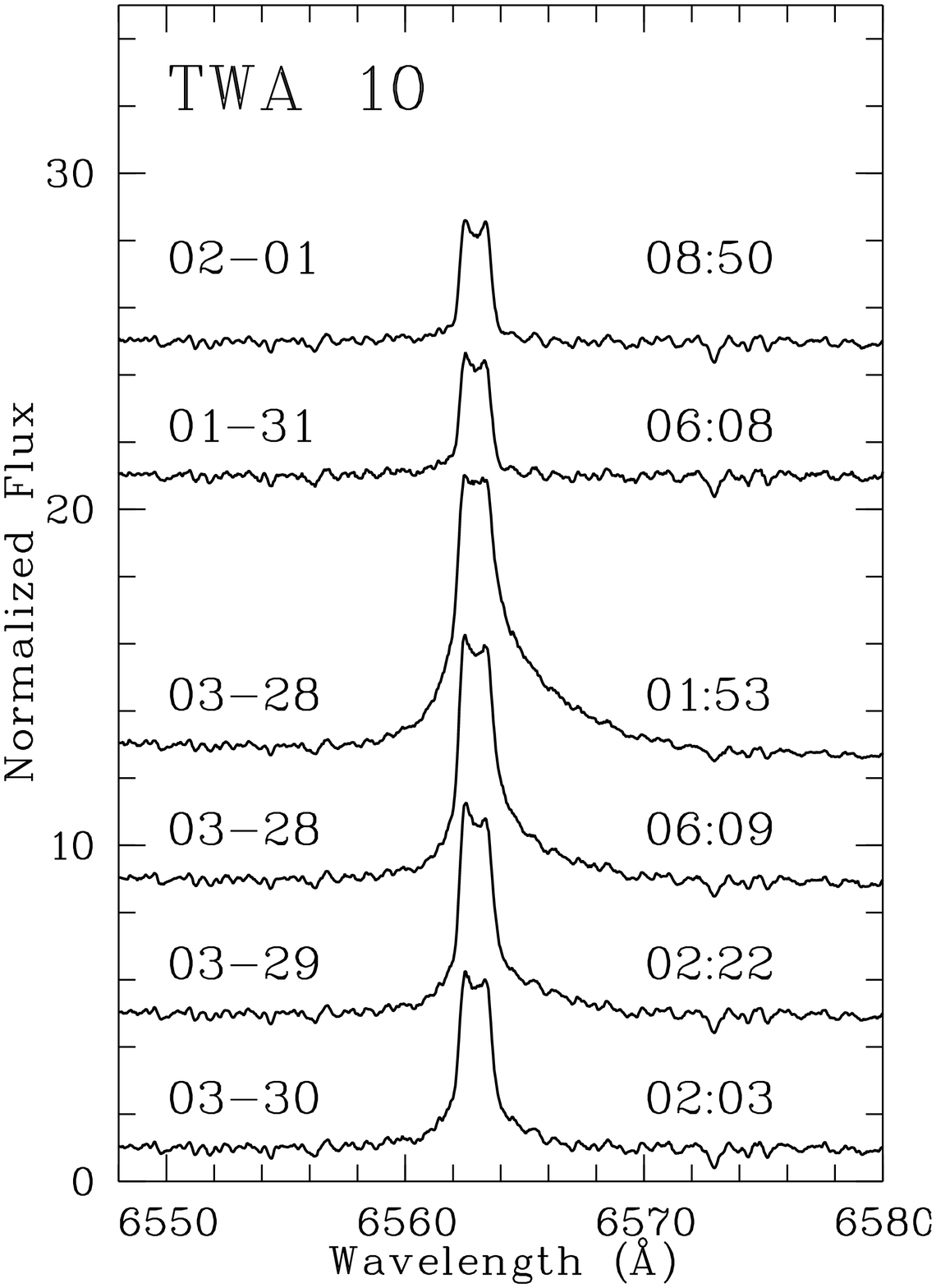}
\caption{H$\alpha$ time series for selected stars in the TW association in chronological order 
(from top to bottom) with UT date and time. Profiles from the different epochs have been shifted 
by arbitrary units for clarity. All profiles are normalised to the continuum.\label{hats2}} 
\end{center}
\end{figure}

As can be seen in Fig. \ref{fig1}, the objects with evidence of accretion 
tend to show particularly strong variability. Out of the seven accretors, five show
H$\alpha$ 10\% width changes by more than $30\,\mathrm{km\,s^{-1}}$. Since the 10\% 
width is correlated with accretion rate \citep{ntm04}, this indicates that the accretion 
flow in these young stars is often rather unsteady and/or clumpy. Whereas the
H$\alpha$ width is variable in many cases, the shape of the line is more or less 
constant for most objects. Some objects, however, show distinct profile changes, probably
due to accretion, binarity, or chromospheric activity. In Figs. \ref{hats1} and \ref{hats2}, 
we present the time series of the H$\alpha$ profiles for some particularly interesting 
objects, including most of the accretors, which we discuss below. 
                                                                                                       
Two objects, {\bf TWA 10} (Fig. \ref{hats2}) and {\bf {\boldmath$\eta$} Cha 5} (Fig. 
\ref{hats1}), show burst-like events in our time series. In both cases, H$\alpha$ 
is clearly much stronger and more asymmetric in one spectrum. For TWA 10, we                           
see gradually decreasing H$\alpha$ intensity after this burst event, with 
the quiescent level reached after about 2 days. A similar trend is seen
in the He\,I (6678\,\AA) emission line, which is a good indicator of accretion, but
occasionally also appears during flare events (see Sect. \ref{accsig}). Additionally,
the H$\alpha$ line appears to be strongly asymmetric during the burst. This hints 
to a link between the burst and ongoing accretion rate changes (see Sect. \ref{accsig}), 
but the lack of mid-infrared emission precludes the existence of a disk for this object 
\citep{lsw05}. Thus, the spectroscopic behaviour is more likely due to a chromospheric 
flare event. For $\eta$ Cha 5, we have less information, because the profile change is 
only apparent in one spectrum. The origin of this event may be a sudden burst of accretion 
or a chromospheric flare. \citet{llm04} observe a broad H$\alpha$ profile in their 
(single) spectrum indicating that sporadic accretion events may occur in this
object. 

The H$\alpha$ time series of {\bf {\boldmath$\eta$} Cha 7} (Fig. \ref{hats1}) shows 
clear signs of binarity (see Sec. \ref{accsig}), even with our sparse time sampling. 
The profile appears to have two components that move relative to each other. The 
two objects with the most complex profiles are {\bf {\boldmath$\eta$} Cha 9} and 
{\bf 11}. Both objects have been identified as accretors in Sect. \ref{accsig}, 
and $\eta$ Cha 9 has a known close companion \citep{kp02}, whereas there is no 
indication of binarity for $\eta$ Cha 11. The H$\alpha$ line 
of $\eta$ Cha 9 (Fig. \ref{hats1}) shows two peaks with a narrow gap in most spectra, where 
the substructure of the red peak changes on timescales of hours. These changes might
be explained by the combined effects of accretion, wind, and binarity. The profile of 
$\eta$ Cha 11 (Fig. \ref{hats1}) shows similar structure in some spectra, but it is highly 
variable in all components. For this object, there is clear evidence for ongoing 
accretion from the H$\alpha$ line profile, because we see a highly red-shifted, 
broad absorption feature, i.e., the line has an inverse P Cygni profile, which is often 
observed in accreting T Tauri stars \citep{r96} and a clear indication of 
infalling material. The complexity of the profile probably indicates strong 
changes in both accretion and wind, as has been found for younger 
T Tauri stars \cite[e.g.][]{jb95,abh05}.

The two well-known accretors in TWA, {\bf TW Hydrae} and {\bf TWA 3A}, show very broad 
and clearly asymmetric profiles (see Fig. \ref{hats2}). Red-shifted absorption features,
most likely due to infalling material, are visible as well as blue-shifted profile
structures, which may be attributed to outflows. The profile changes are, however, not
as dramatic as for the objects discussed above.

The H$\alpha$ variability of {\bf TWA 5A} was of particular interest to us because it 
had been suspected to be an accretor previously \citep{mjb03} and an AO 
resolved close binary \citep{bjn03}, with roughly equal mass components and a separation 
of only 54\,mas, corresponding to a projected separation of 3\,AU at a distance of 
55\,pc. Additionally, previous high-resolution spectra had hinted that one of the
AO components may be a spectroscopic binary as well \citep{t03}. Therefore TWA 5A was
monitored in a more extended time series than most of our other targets; in total we
obtained 22 spectra, 17 during the 2005 March observing run. 

The intensity of H$\alpha$ is strongly variable in our time series. The time series
of the EW shows a strong burst in the night of the 28th of March, which has strong 
influence on the average EW for this object. This burst event coincides with the
appearance of He\,I and Ca\,II in the spectra (see Sec. \ref{accsig}), which are 
not or only weakly detected in the remaining spectra. Only during this event, the EW 
exceeds significantly the adopted threshold between accretors and non-accretors, whereas 
the 10\% width is always in the regime normally attributed to accretors (see Sect.
\ref{accsig}). However, accretion is unlikely to be the origin for the emission, as argued 
in Sect. \ref{accsig}, because of the non-detection of a disk excess out to the far-infrared. 
Thus, the profile shape and variability are probably caused by the combined effects of 
chromospheric activity, rotation, and binarity. Indeed, the spectroscopic signature of 
the burst bears similarities to previously observed chromospheric flares \citep{msc98,ma01}. 
Thus, we attribute this event to a magnetically induced flare.

\begin{figure}[t]
\begin{center}
\includegraphics[width=10cm,angle=0]{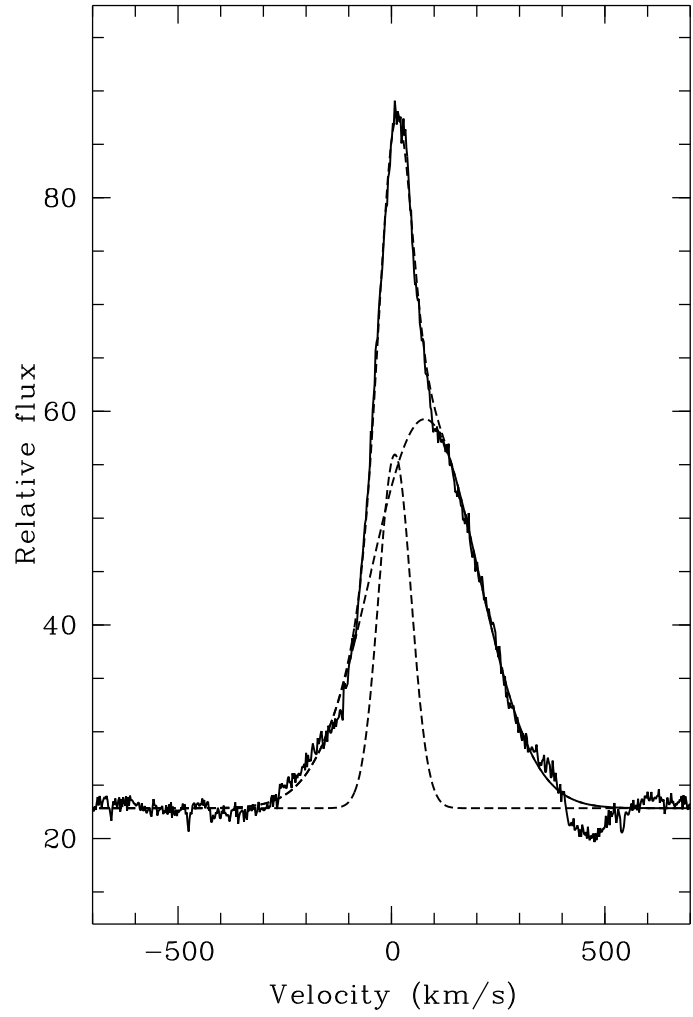}
\caption{The average H$\alpha$ line profile for TWA 5A from our time series. The solid line
is the observed profile, whereas the dashed lines show results from the decomposition
process (narrow component, broad component, and their sum). See discussion in Sect.
\ref{variab}. \label{twa5a_decomp}} 
\end{center}
\end{figure}

The general shape of the H$\alpha$ profiles are very similar in all our spectra: 
it appears to be composed by a broad and a narrow component, where both 
components clearly change in intensity and relative position (see Fig. \ref{hats2}). 
We therefore approximated the H$\alpha$ of TWA 5A with two Gaussians to disentangle the two 
components. In all spectra, the profile composed of two Gaussians fit the observed 
profile almost perfectly, as can be seen in Fig. \ref{twa5a_decomp}. The typical 
$\sigma$ is $130\,\mathrm{km\,s^{-1}}$ for the broad component and 
$40\,\mathrm{km\,s^{-1}}$ for the narrow one, corresponding to EW of 10-15\,\AA\, 
and $\sim 5$\,\AA, respectively.

What is the origin of broad and narrow components? Perhaps the most likely interpretation 
is that we observe emission from two strongly magnetically active stars in this binary system.
In this case, the very different linewidths in the two components might be explained by 
differences in rotational velocities. If one of the AO-resolved stars is again a spectroscopic
binary, where both stars are H$\alpha$ active, this would also allow for a larger 
linewidth and thus provide an explanation for the broad component. Another possible 
explanation for the line profile is that both components arise from one particularly 
active star in the system, which exhibits two active regions with different characteristics 
in the chromosphere. 

A more detailed investigation of the complex effects of activity, binarity, and rotation on 
the H$\alpha$ feature requires prior knowledge on the properties of the two (or possibly three)
stars in the system. Thus, a sophisticated analysis of the photospheric lines in the composite
TWA 5A spectrum is necessary, to determine rotational velocities for both stars and
to investigate mass and orbital period of a potential third body in the system.

\section{Conclusions}

We have carried out an extensive investigation of disk accretion and related
phenomena in 100 young stars in four nearby associations that span 6-30 Myr
in age. Our study is based on $\sim$650 high-resolution, high
signal-to-noise multi-epoch optical spectra of these targets. We find that disk
accretion ceases or dips below measurable levels by $\sim$10 Myr in the
vast majority of low-mass stars. This result provides an indirect constraint
on the timescale for gas dissipation in inner disks and, thus, gas giant
planet formation. We find that all accretors in our sample are slow rotators,
whereas non-accretors are either slow or fast rotators. This may indicate that
stellar rotation is still braked by the disk at 8\,Myr, but the small number
of accretors at these ages hampers a more detailed investigation of
the rotation-accretion connection. Accretors also often exhibit significant time 
variability in their emission lines, providing evidence for clumpy accretion 
flows and/or episodic accretion events.

\acknowledgments
We thank the Las Campanas Observatory staff for their outstanding assistance. 
This research was supported by NSERC grants to RJ and MHvK as well as 
University of Toronto startup funds and an SAO subcontract for the Keck Nuller 
project to RJ. JC was supported in part by an NSERC Undergraduate Student Research 
Award.

\clearpage
\newpage

\clearpage
\newpage

\clearpage
\newpage

\clearpage
\newpage

\clearpage
\newpage

\begin{deluxetable}{ll}
\tabletypesize{\scriptsize}
\rotate
\tablecaption{Observation Record \label{obsrec}}
\tablewidth{0pt}
\tablehead{
\colhead{Date (UT)} & \colhead{Targets (in brackets: no. of observations if $>1$)}
}
\startdata

2004-12-04 & $\eta$ Cha 1, $\eta$ Cha 11, $\eta$ Cha 7, TWA 19A, TWA 5A, TWA 9A, AU Mic (3), AO Men (2), $\beta$ Pic, GJ3305 (2), HIP 23309 (2). \\

2004-12-05 & $\eta$ Cha 4, $\eta$ Cha 5, $\eta$ Cha 7, $\eta$ Cha 10, $\eta$ Cha 11, $\eta$ Cha 12, AO Men (2), GJ3305 (2), HIP 23309 (2), TW Hya (2), TWA 5A, TWA 9A, TWA 19A. \\

2004-12-06 & $\eta$ Cha 3, $\eta$ Cha 4, $\eta$ Cha 6, $\eta$ Cha 7, $\eta$ Cha 9, $\eta$ Cha 11, $\eta$ Cha 13, TW Hya (2),  TWA 5A, TWA 6, TWA 7, TWA 8A, TWA 9A, TWA 19A, AO Men (2), \\ & AU Mic (5), $\beta$ Pic (2), GJ3305 (2), HIP 23309 (2). \\

2005-01-29 & $\eta$ Cha 1 (2), $\eta$ Cha 10 (2), $\eta$ Cha 11, $\eta$ Cha 12 (2), $\eta$ Cha 13, $\eta$ Cha 3, $\eta$ Cha 4 (2), $\eta$ Cha 5, $\eta$ Cha 6 (2), $\eta$ Cha 9 (2), TW Hya, TWA 2A, TWA 3A, \\ & TWA 3B, TWA 4A, TWA 6, V343Nor. \\

2005-01-30 & $\eta$ Cha1, $\eta$ Cha 3 (2), $\eta$ Cha4, $\eta$ Cha 5 (2), $\eta$ Cha6, $\eta$ Cha10, $\eta$ Cha12, $\eta$ Cha13 (2), TWA 2A, TWA 3A, TWA 3B, TWA 4A, TWA 5A, TWA 6, TWA 7, \\  & TWA 8A, TWA 9A, TWA 9B, TWA 11A, TWA 11B, TWA 13A, TWA 13B, TWA 19A, TWA 19B, AO Men, $\beta$ Pic (2), HD 35850, HIP 23309, V343Nor, \\ & CD-53544, CD-60416, CPD-64120, GSC8056-0482,GSC8491-1194, GSC8497-0995, HD 13183, HD 13246, HD 8558, HD 9054, HIP 16853, HIP 21632, \\ & HIP 22295, HIP 30030, HIP 30034, HIP 32235, HIP 33737, HIP 9141, TYC5882-1169, TYC7600-0516. \\

2005-01-31 & $\eta$ Cha 1, $\eta$ Cha 4, $\eta$ Cha 11 (2), $\eta$ Cha 12, TWA 2A, TWA 4A, TWA 5A, TWA 6, TWA 7, TWA 9A, TWA 10, TWA 12, TWA 13A, TWA 13B, TWA 14, \\ & TWA 16, TWA 17, TWA 18, TWA 19A, TWA 19B, $\beta$  Pic, AO Men, HD 35850, HIP 23309, CD-53544, CD-60416, CPD-64120, GSC8056-0482, GSC8491-1194, \\ & GSC8497-0995, HD 13183, HD 13246, HD 8558, HD 9054, HIP 16853, HIP 21632, HIP 30030, HIP 32235, HIP 33737, HIP 9141, TYC5882-1169, TYC7600-0516. \\

2005-02-01 & $\eta$ Cha 1, $\eta$ Cha 3, $\eta$ Cha 4, $\eta$ Cha 5, $\eta$ Cha 6 (2), $\eta$ Cha 9, $\eta$ Cha 10, $\eta$ Cha 12 (2), $\eta$ Cha 13, TW Hya (2), TWA 2A, TWA 4A, TWA 5A, TWA 6, TWA 7, \\ & TWA 8A, TWA 8B, TWA 9A, TWA 9B, TWA 10, TWA 12, TWA 13A, TWA 13B, TWA 14, TWA 17, TWA 18, TWA 19A, TWA 19B, $\beta$  Pic, GJ3305, \\ & HD 35850, HIP 23309, V343Nor, CD-53544, CD-60416, CPD-64120, GSC8056-0482,GSC8491-1194, GSC8497-0995, HD 13183, HD 13246, HD 8558, \\ & HD 9054, HIP 16853, HIP 21632, HIP 22295, HIP 30030, HIP 30034, HIP 32235, HIP 33737, HIP 9141, TYC5882-1169, TYC7600-0516.\\

2005-03-26 & TWA 3A, TWA 3B, HIP 23418. \\

2005-03-27 & $\eta$ Cha 1, $\eta$ Cha 3, $\eta$ Cha 4, $\eta$ Cha 13, TW Hya (2), TWA  2A (2), TWA  5A (4), TWA  6 (2), TWA  7 (2), TWA  8A (2), TWA  8B (2), TWA  15A, TWA  15B, \\ & TWA  16, TWA  20, TWA  21, TWA  22, TWA  23, TWA  24, TWA  25,  $\beta$ Pic (2), AO Men, AU Mic (4), BD-17d6128, CD-64d1208, GJ 3305, GJ799A, GJ799B, \\ & HD  164249, HD 181327, HD 199143,
 HD 35850 (2), HIP 23309, HIP 23418, HR7329B, PZTel, HIP 105388, HIP 105404, HIP 107345, HIP 107947, HIP 108422, \\ & HIP 16853 (2), HIP 21632, HIP 22295, HIP 30030, HIP 30034, HIP 32235, HIP 33737, TYC5882-1169, TYC7065-0879 (2), TYC7600-0516. \\

2005-03-28 & $\eta$ Cha 5, $\eta$ Cha 6, $\eta$ Cha 9, $\eta$ Cha 10, $\eta$ Cha 11, $\eta$ Cha 12, TWA  5A (5), TWA  9A (2), TWA  9B (2), TWA  10 (2), TWA  11A (2), TWA  11B (2), TWA  12 (2), \\ & TWA  13A (2),  TWA  13B (2), TWA  14 (2), TWA  16 (2), TWA  17 (2), TWA  18 (2), TWA  19A (2), TWA  19B (2), TWA  20, TWA  21 (2), TWA  22 (2), TWA  23, \\ & TWA  24 (2), TWA  25, AU Mic (3), BD-17d6128 (2), $\beta$ Pic, CD-64d1208, GJ799A, GJ799B, HD 164249, HD 181327, HD 199143, PZTel, V343Nor, HIP 105388, \\ & HIP 105404, HIP 107345, HIP 107947, HIP 108422, HIP 21632, HIP 22295, HIP 30030, HIP 30034, HIP 32235, HIP 33737, TYC5882-1169, TYC7600-0516. \\

2005-03-29 & TW Hya, TWA  2A, TWA  3A, TWA  4A, TWA  5A (4), TWA  6, TWA  7, TWA  8A, TWA  8B, TWA  9A, TWA  9B, TWA  10, TWA  11A, TWA  11B, TWA  12, \\ & TWA  13A, TWA  13B, TWA  14, TWA  15A (2), TWA  15B (2), TWA  17, TWA  18, TWA  19A, TWA  19B, TWA  20 (2), TWA  21 (2), TWA  22 (2), TWA  23 (2), \\ & TWA 24, TWA 25 (2), AU Mic (7), BD-17d6128, $\beta$ Pic (2), CD-64d1208, GJ799A, GJ799B, HD 164249, HD 181327, HD 199143, HIP 23418, PZTel, HIP 105388, \\ & HIP 105404, HIP 107345, HIP 107947, HIP 108422. \\

2005-03-30 & TW Hya, TWA 2A, TWA 3A, TWA 3B, TWA 4A, TWA 5A (4), TWA 6, TWA 7, TWA 8A, TWA 8B, TWA 9A, TWA 9B, TWA 10, TWA 11A (13), TWA 11B, \\ & TWA 12, TWA 13A, TWA 13B, TWA 14, TWA 15A (2), TWA 15B (2), TWA 16, TWA 18, TWA 19A, TWA 19B, TWA 20 (2), TWA 21, TWA 22, TWA 23 (2), \\ &  TWA 24, TWA 25 (2), AU Mic (2), BD-17d6128, CD-64d1208, GJ799A, GJ799B, HD 164249, HD 181327, HD 199143, HIP 23418, PZ Tel, HIP 105388, HIP 105404, \\ & HIP 107345, HIP 107947, HIP 108422.  \\

2005-07-19 & TWA 2A, TWA 3A, TWA 3B, TWA 4A, TWA 5A, TWA 6, TWA 7, TWA 8A, TWA 8B, TWA 9A, TWA 9B, TWA 12, TWA 13A, TWA 13B, \\ & TWA 14, TWA 16, TWA 17, TWA 18, TWA 19A, TWA 19B, TWA 21, TWA 22, TW Hya, AU Mic (2), BD-17d6128 (2), $\beta$ Pic (5), CD-64d1208 (2), \\ & GJ799A (2), GJ799B (2), HD 1555A (2), HD 164249 (2), HD 181327 (2), HD 199143 (2), HD 35850, HIP 112312 (2), HIP 112312B (2), HIP 12545 (2), \\ & HIP 23418, PZ Tel (2), V343Nor, CD-53544, CD-60416 (2), CPD-64120 (2), GSC8056-0482, GSC8491-1194, GSC8497-0995, HD 13183 (2), HD 13246 (2), \\ & HD 8558 (2), HD 9054 (2), HIP 105388 (2), HIP 105404 (2), HIP 107345 (2), HIP 107947 (2), HIP 108422 (2), HIP 1113 (2), HIP 116748B (2), \\ & HIP 116748A (2), HIP 1481 (2), HIP 16853, HIP 1910 (2), HIP 1993 (2), HIP 21632, HIP 22295, HIP 2729 (2), HIP 3556 (2), HIP 490 (2), \\ & HIP 9141 (2), TYC5882-1169, TYC7600-0516.  \\

\enddata
\end{deluxetable}

\clearpage
\newpage

\begin{deluxetable}{llcllccccclllllcccccl}
\tabletypesize{\scriptsize}
\rotate
\tablecaption{Summary of Results: $\eta$ Cha \label{res1}}
\tablewidth{0pt}
\tablehead{
\colhead{Object} & \colhead{Other Name} & \colhead{No. of spectra} & \colhead{Sp. type} & \colhead{EW} & 
\colhead{EW $\sigma$\tablenotemark{a}} & \colhead{10\% width} & \colhead{10\% width $\sigma$\tablenotemark{a}} & \colhead{v$\sin i$\tablenotemark{d}} &\colhead{Comments} \\
\colhead{} & \colhead{} &  \colhead{} & \colhead{} & \colhead{(\AA)} & \colhead{(\AA)} 
& \colhead{($\mathrm{km\,s^{-1}}$)} & \colhead{($\mathrm{km\,s^{-1}}$)} & \colhead{($\mathrm{km\,s^{-1}}$)} & \colhead{}}
\tablecolumns{10}
\startdata

$\eta$ \ Cha 1  &  RECX 1            & 7 & K4\tablenotemark{b}    & -1.2  & 0.3 &  103 & 13  & 20.9 \\
$\eta$ \ Cha 3  &  RECX 3            & 6 & M3.25\tablenotemark{c} & -2.0  & 0.2 &  116  & 8  & 10.5 \\
$\eta$ \ Cha 4  &   RECX 4           & 8 & K7\tablenotemark{b}     & -3.4  & 0.6 &  147 & 13 &   6.0 \\
$\eta$ \ Cha 5  & RECX 5             & 6 & M4\tablenotemark{c}     & -8.6  & 4.3 &  194 & 66 &   8.8 & \\
$\eta$ \ Cha 6  &   RECX 6           & 7 & M2\tablenotemark{b}     & -5.0  & 0.4  & 145 & 27 &  20.9 \\
$\eta$ \ Cha 7  &  RECX 7            & 3 & K6\tablenotemark{c}     & -1.0  & 0.3  & 291 & 70  &  29.1 & spectroscopic binary \\
$\eta$ \ Cha 9  &  RECX 9            & 6 & M4.5\tablenotemark{c}   & -11.7  & 2.7 &  389 & 54  &  20.8 & He\,I\,(6678\,\AA)\\
$\eta$ \ Cha 10 &  RECX 10           & 6 & K7\tablenotemark{b}     & -1.2  & 0.3  & 103 & 13 & $\mathrm{<}$ 5.0 \\
$\eta$ \ Cha 11 & RECX 11            & 7 & K4\tablenotemark{b}     & -3.9  & 2.1 &  345 & 86 &  16.4 & \\
$\eta$ \ Cha 12 &  RECX 12           & 8 & M2\tablenotemark{b}     & -5.7  & 0.5  & 154  & 7 &  15.1 \\
$\eta$ \ Cha 13 &  ECHA J0843.3-7905 & 6 & M2\tablenotemark{b}      & -110.8 & 13.7  & 435 & 26  &  14.5 & He\,I\,(6678\,\AA) \\
\enddata
\tablenotetext{a}{Values for $\sigma$ are the scatter in our multi-epoch spectra, not the measurement uncertainty.
Errors in EW are 0.2\,\AA, in 10\% width 5\,km\,s$^{-1}$}
\tablenotetext{b}{\citet{zs04}}
\tablenotetext{c}{\citet{ls04}}
\tablenotetext{d}{The values for spectroscopic binaries (see last column) might be affected by line blending.}
\end{deluxetable}

\clearpage
\newpage

\begin{deluxetable}{llcllccccclllllcccccl}
\tabletypesize{\scriptsize}
\rotate
\tablecaption{Summary of Results: TW Hydrae \label{res2}}
\tablewidth{0pt}
\tablehead{
\colhead{Object} & \colhead{Other Name} & \colhead{No. of spectra} & \colhead{Sp. type} & \colhead{EW} 
& \colhead{EW $\sigma$\tablenotemark{a}} & \colhead{10\% width} & \colhead{10\% width $\sigma$\tablenotemark{a}} & \colhead{v$\sin i$\tablenotemark{f}} &  \colhead{Comments} \\
\colhead{} & \colhead{} & \colhead{} &  \colhead{} & \colhead{(\AA)} & \colhead{(\AA)} 
& \colhead{($\mathrm{km\,s^{-1}}$)} & \colhead{($\mathrm{km\,s^{-1}}$)} & \colhead{($\mathrm{km\,s^{-1}}$)} & \colhead{}}
\tablecolumns{10}
\startdata

TW Hya      &   TWA 1           & 12 & K7\tablenotemark{b}   & -172.5  & 44.7 & 425  & 35   & 10.6  & He\,I\,(6678\,\AA)\\
TWA 2A      &   CD -29 8887A    &  9 & M2e\tablenotemark{c}  & -1.8   & 0.2 &  134  & 17   & 12.8 \\
TWA 3A      &   Hen 3-600A      & 5 & M3e\tablenotemark{c}   & -22.6   & 2.9  & 262   & 4   & 11.6 & spectroscopic binary, He\,I\,(6678\,\AA)\tablenotemark{d}\\
TWA 3B      &   Hen 3-600B      & 4 & M3.5\tablenotemark{c} &  -6.1   & 0.9  & 153  & 13   & 12.2 & \\
TWA 5A      &   CD -33 7795A    & 24 & M1.5\tablenotemark{c} & -11.5   & 4.9  & 360  & 54   & 59.0 & He\,I\,(6678\,\AA) \\
TWA 6       &   TYC 7183-1477   & 10 & K7\tablenotemark{b}   & -3.4   & 0.4  & 409  & 42   & 79.5 & spectroscopic binary\\
TWA 7       &   TYC 7190-2111   & 9  & M1\tablenotemark{b}   &  -5.8   & 0.9  & 109   & 8  & $\mathrm{<}$ 5.0 \\
TWA 8A      &   \nodata         & 8  & M2\tablenotemark{c}   &   -8.0   & 1.4  & 133  & 29   & $\mathrm{<}$ 5.0  \\
TWA 8B      &   \nodata         & 6  & \nodata &  -13.3     & 1.8  & 119   & 10   & 11.2\\
TWA 9A      &   CD -36 7429A    & 11 & K5\tablenotemark{c}   &  -2.1   & 0.5  & 135  & 21   & 11.3 \\
TWA 9B      &   CD -36 7429B    & 7 & M1\tablenotemark{c}   & -4.3 & 0.6 & 108 & 12 & 8.4\\
TWA 10      &   GSC 07766-00743 & 6 & M2.5\tablenotemark{b} & -13.6   & 9.6  & 199  & 97    & 6.3 & He\,I\,(6678\,\AA)\\
TWA 11B     &   HR 4796B        & 5 & M2.5\tablenotemark{c} & -3.5   & 0.6 &  116  & 10   & 12.1 \\
TWA 12      &   RX J1121.1-3845 & 7 & M2\tablenotemark{c}   & -4.8   & 0.9  & 136  & 28   & 16.2 \\
TWA 13A     &   RX J1121.3-3447N& 8 & M1e\tablenotemark{c}  & -3.0   & 0.7  & 129  & 24   & 10.5\\
TWA 13B     &   RX J1121.3-3447N& 8 & M2e\tablenotemark{c}  & -3.0   & 0.7  & 134  & 26   & 10.3\\
TWA 14      &   \nodata         & 7 & M0\tablenotemark{c}   & -10.7   & 7.8 &  291  & 61   & 43.1 & spectroscopic binary \\
TWA 15A     &   \nodata         & 5 & M1.15\tablenotemark{c} & -8.8   & 0.5  & 156   & 2   & 21.3\\
TWA 15B     &   \nodata         & 5 & M2\tablenotemark{c}   & -8.6   & 1.4  & 193  & 16   & 32.3\\
TWA 16      &   \nodata         & 6 & M1.5\tablenotemark{c} & -4.0   & 0.8  & 119  & 12    & 7.9\\
TWA 17      &   \nodata         & 6 & K5\tablenotemark{c}   & -3.2   & 0.4  & 290  & 25   & 49.7 & spectroscopic binary \\
TWA 18      &   \nodata         & 7 & M0.5\tablenotemark{c} & -3.3   & 0.4  & 179  & 11   & 24.1\\
TWA 19B     &   HD 102458B      & 8 & K7\tablenotemark{c}   &  -2.2   & 0.4 &  239  & 45   & 48.7 & spectroscopic binary\\
TWA 20      &   A2 146          & 6 & M2\tablenotemark{c}   &  -3.1   & 0.2 &  273  & 28 & \tablenotemark{e} & spectroscopic binary \\
TWA 22      &   SSS 1017-5354   & 7 & M5\tablenotemark{b}   & -11.5  & 2.0 &  111   & 10    & 9.7\\
TWA 23      &   SSS 1207-3247   & 6 & M1\tablenotemark{b}   & -2.4   & 0.2  & 126   & 3   & 14.8 \\
TWA 24      &   GSC8644-0802    & 5 & K3\tablenotemark{b}  & -0.3      & 0.2  & 137  & 29   & 13.0 \\
TWA 25      &   TYC 7760-0283   & 6 & M0\tablenotemark{b}   &  -2.4   & 0.5  & 128  & 10   & 11.8 \\
\enddata
\tablenotetext{a}{Values for $\sigma$ are the scatter in our multi-epoch spectra, not the measurement uncertainty.
Errors in EW are 0.2\,\AA, in 10\% width 5\,km\,s$^{-1}$}
\tablenotetext{b}{\citet{zs04}}
\tablenotetext{c}{\citet{dp04}}
\tablenotetext{d}{For all measurements, only spectra were used where we see no doubled lines. Linewidths 
should therefore not be affected by binarity.}
\tablenotetext{e}{TWA 20 is a spectroscopic binary; it was not possible to obtain a $v\sin i$ fit with acceptable
reliability.}
\tablenotetext{f}{The values for spectroscopic binaries (see last column) might be affected by line blending.}
\end{deluxetable}

\clearpage
\newpage

\begin{deluxetable}{llcllccccclllllcccccl}
\tabletypesize{\scriptsize}
\rotate
\tablecaption{Summary of Results: $\beta$ Pic moving group\label{res3}}
\tablewidth{0pt}
\tablehead{
\colhead{Object} & \colhead{Other Name} & \colhead{No. of spectra} & \colhead{Sp. type} & \colhead{EW} & \colhead{EW $\sigma$\tablenotemark{a}} 
& \colhead{10\% width} & \colhead{10\% width $\sigma$\tablenotemark{a}} & \colhead{v$\sin i$\tablenotemark{d}} &  \colhead{Comments} \\
\colhead{} & \colhead{} &  \colhead{} &  \colhead{} & \colhead{(\AA)} & \colhead{(\AA)} 
& \colhead{($\mathrm{km\,s^{-1}}$)} & \colhead{($\mathrm{km\,s^{-1}}$)} & \colhead{($\mathrm{km\,s^{-1}}$)} & \colhead{}}
\tablecolumns{10}
\startdata
AO Men       &  HIP29964     & 9 & K6/7\tablenotemark{b}  &   -0.6 & 0.1 & 122 & 15 & 16.0 \\
Au Mic       &  HIP102409   &  26 & \nodata    &     -2.3   & 0.8  & 134  & 52 & 8.5 \\
BD-17 6128  &  HD 358623    & 7 & K7/M0\tablenotemark{c}  &   -0.5   & 0.2 &  118  & 13 & 14.6\\
CD-64 1208  &  CPD-64 3950 & 6 & K7\tablenotemark{b}     &    -2.9   & 0.5 &  494  & 23 & 102.7 & spectroscopic binary  \\
GJ3305   &   \nodata       & 8  &  M0.5\tablenotemark{b}      &     -2.2   & 0.3 &  128  & 15 & 5.3 \\
GJ799A   &   \nodata       & 6  &  M4.5e\tablenotemark{b}     &    -10.5   & 1.4  & 96   & 3 & 10.6 \\
GJ799B   &  \nodata        & 6  &  M4.5e\tablenotemark{b}    &    -8.9   & 0.8 &  138 & 13 & 17.0 \\
HIP 12545 & BD+05 378 & 2 & M0\tablenotemark{b} &  -0.6 & 0.0 & 95 & 5  & 9.3\\
HIP 23309    & CD-57 1054  & 10 & M0.5\tablenotemark{b}      &     -0.8   & 0.2 &  108  & 13  &  5.8 \\
HIP 23418A   &  GJ 3322A & & M3\tablenotemark{b}    &      -6.6 & \nodata & 108 & \nodata & 7.7  \\
HIP 23418B  & GJ 3322B & & \nodata & -6.1  & \nodata & 116.6 & \nodata  & 21.0 \\
HIP 112312A & V* WW PsA & 2 & M4e\tablenotemark{b}  & -6.6   & 0.1 &  105   & 4 & 14.0 \\
HIP 112312B & \nodata         & 2 & M4.5e\tablenotemark{b} &  -8.2   & 0.5 &  137   & 4 & 24.3\\
\enddata
\tablenotetext{a}{Values for $\sigma$ are the scatter in our multi-epoch spectra, not the measurement uncertainty.
Errors in EW are 0.2\,\AA, in 10\% width 5\,km\,s$^{-1}$}
\tablenotetext{b}{\citet{zs04}}
\tablenotetext{c}{\citet{dp04}}
\tablenotetext{d}{The values for spectroscopic binaries (see last column) might be affected by line blending.}
\end{deluxetable}

\clearpage
\newpage

\begin{deluxetable}{llcllccccclllllcccccl}
\tabletypesize{\scriptsize}
\rotate
\tablecaption{Summary of Results: Tuc-Hor \label{res4}}
\tablewidth{0pt}
\tablehead{
\colhead{Object} & \colhead{Other Name} & \colhead{No. of spectra} & \colhead{Sp. type} & \colhead{EW} 
& \colhead{EW $\sigma$\tablenotemark{a}} & \colhead{10\% width} & \colhead{10\% width $\sigma$\tablenotemark{a}} & \colhead{vsini} &  \colhead{Comments} \\
\colhead{} & \colhead{} & \colhead{} &  \colhead{} & \colhead{(\AA)} & \colhead{(\AA)} 
& \colhead{($\mathrm{km\,s^{-1}}$)} & \colhead{($\mathrm{km\,s^{-1}}$)} & \colhead{($\mathrm{km\,s^{-1}}$)} & \colhead{}}
\tablecolumns{10}
\startdata

CD-53544    &  TYC 8491- 656-1 & 4 & K6Ve\tablenotemark{b} & -1.4   & 0.2 &  403  & 48   & 82.2 & fast rotator\\
CD-60416    &  TYC 8489- 1155-1 & 5 & K3/4\tablenotemark{b} &  -0.5   & 0.0  & 110  & 19   & 10.1\\
CPD-64120   &  TYC 8852- 264-1 & 5 & K1Ve\tablenotemark{b} &  -0.2   & 0.1 &  103  & 15   & 30.2\\
GSC8056-0482 & \nodata  & 4 & M3Ve\tablenotemark{b} & -5.3   & 0.4  & 177  & 26   & 34.2\\
GSC8491-1194 &  \nodata & 4 & M3Ve\tablenotemark{b} & -4.1   & 0.4  & 99   & 1   & 12.8\\
GSC8497-0995 & \nodata  & 4 & K6Ve\tablenotemark{b} &  -0.6   & 0.2  & 96  & 12    & 6.6\\
HIP 1910   & BPM 1699 & 2 & M1\tablenotemark{b} &  -1.6   & 0.1  & 132   & 2   & 19.0\\
HIP 1993  & GSC 08841-00145 & 2 & M1\tablenotemark{b} &  -1.0   & 0.1  & 97   & 8    & 7.1 \\
HIP 2729  & HD 3221 & 2 & K5V\tablenotemark{b} & -0.7   & 0.3  & 467  & 28  & 127.5 & fast rotator\\
HIP 3556  & GJ 3054 & 2 & M3\tablenotemark{b} &  -0.8   & 0.0  & 107  & 12 & $\mathrm{<}$ 5.0  \\
HIP 107345   & BPM 14269 & 6 & M1\tablenotemark{b} &  -1.4   & 0.2  & 114  & 13  & $\mathrm{<}$ 5.0  \\
\enddata
\tablenotetext{a}{Values for $\sigma$ are the scatter in our multi-epoch spectra, not the measurement uncertainty.
Errors in EW are 0.2\,\AA, in 10\% width 5\,km\,s$^{-1}$}
\tablenotetext{b}{\citet{zs04}}
\end{deluxetable}

\end{document}